\documentclass[12pt]{article}

\usepackage{graphicx,amssymb,url,mathrsfs, amsmath}
\usepackage{eucal}
\usepackage{wrapfig}
\usepackage{boxedminipage}
\usepackage{setspace}
\usepackage{subfigure}
\usepackage{amsxtra,amstext,latexsym,dsfont}

\def\a  {\alpha}       \def\b  {\beta}         \def\g  {\gamma}
       \def\d  {\delta}        \def\D  {\Delta}
\def\e  {\epsilon}        \def\k  {\kappa}
\def\l  {\lambda}             \def\m  {\mu}
\def\n  {\nu}          \def\s  {\sigma}        
\def\t  {\tau}                 
   \def\w  {\omega}

\newcommand{\cala}{\mbox{${\cal A}$}} 
 
 \newcommand{\calf}{\mbox{${\cal F}$}}

\newcommand{\calo}{\mbox{${\cal O}$}} 
\newcommand{\calq}{\mbox{${\cal Q}$}}

\newcommand{\scra}{\mbox{${\mathscr A}$}}

\def\IR{{\hbox{{\rm I}\kern-.2em\hbox{\rm R}}}}
\def\IB{{\hbox{{\rm I}\kern-.2em\hbox{\rm B}}}}
\def\IN{{\hbox{{\rm I}\kern-.2em\hbox{\rm N}}}}
\def\IC{\,\,{\hbox{{\rm I}\kern-.59em\hbox{\bf C}}}}
\def\IZ{{\hbox{{\rm Z}\kern-.4em\hbox{\rm Z}}}}
\def\IP{{\hbox{{\rm I}\kern-.2em\hbox{\rm P}}}}
\def\IH{{\hbox{{\rm I}\kern-.4em\hbox{\rm H}}}}
\def\ID{{\hbox{{\rm I}\kern-.2em\hbox{\rm D}}}}

\def\be{\begin{equation}}
\def\ee{\end{equation}}
\def\ba{\begin{eqnarray}}
\def\ea{\end{eqnarray}}

\def\half{\frac{1}{2}}

\newcommand{\inv}[1]{\frac{1}{#1}}

\def\ra{\rightarrow}

\newcommand{\ud}{\mbox{${\mathrm{d}}$}}

\def\dell{\partial}

\newcommand{\abs}[1]{\left| #1 \right|}

\def\dg{{\dagger}}
\def\Tr{{\rm tr}\,}

\def\nn{\nonumber}
\def\ea{{\it et al}. }

\def\DeB{\overline{\textrm{D8}}}
\def\DeDeB{\textrm{D8-}\overline{ \textrm{D8}}}

\newcommand{\Mkk}{M_{\rm KK}}
\newcommand{\wt}{\widetilde}
\newcommand{\wh}{\widehat}

\newcommand{\mone}{{{  \mathds{1} }}}

\begin{document}

\begin{titlepage}

\vspace{0.5in}

\begin{center}
{\large \bf Nucleon-Nucleon Potential from Holography} \\
\vspace{10mm}
  Keun-Young Kim and Ismail Zahed\\
\vspace{5mm}
{\it Department of Physics and Astronomy, SUNY Stony-Brook, NY 11794}\\
\end{center}
\begin{abstract}

In the holographic model of QCD, baryons are chiral solitons sourced
by D4 flavor instantons in bulk of size $1/\sqrt{\lambda}$ with
$\lambda=g^2N_c$. Using the ADHM construction we explicit the exact
two-instanton solution in bulk. We use it to construct the core NN
potential to order $N_c/\l$. The core sources meson fields to order
$\sqrt{N_c/\l}$ which are shown to contribute to the NN interaction to
order $N_c/\l$. In holographic QCD, the NN interaction splits into
a small core and a large cloud contribution in line with meson
exchange models. The core part of the interaction is repulsive in
the central, spin and tensor channels for instantons in the regular
gauge. The cloud part of the interaction is dominated by omega exchange
in the central channel, by pion exchange in the tensor channel
and by axial-vector exchange in the spin and tensor channels. Vector
meson exchanges are subdominant in all channels.

\end{abstract}

\end{titlepage}

\renewcommand{\thefootnote}{\arabic{footnote}}
\setcounter{footnote}{0}



\section{Introduction}

Holographic QCD has provided an insightful look to a number of
issues in baryonic physics at strong coupling $\lambda=g^2N_c$
and large number of colors $N_c$~\cite{BRODSKY,RHO,Sakai3,
KSZ,SEITZ,Hashimoto,FF,Baryons2,Baryons3,Finitemu}.
In particular, in \cite{BRODSKY,RHO} baryons
are constructed from a five-dimensional Shrodinger-like equation
whereby the 5th dimension generates mass-like anomalous
dimensions through pertinent boundary conditions. A number of
baryonic properties have followed ranging from baryonic spectra
to form factors \cite{BRODSKY,RHO}.

At large $N_c$ baryons are chiral solitons in QCD. A particularly
interesting framework for discussing this scenario is the
D8-$\DeB$ chiral holographic model recently suggested by Sakai and
Sugimoto~\cite{ Sakai1,Sakai3} (herethrough hQCD). In hQCD
D4 static instantons in bulk source the chiral solitons or
Skyrmions on the boundary. The instantons have a size of order
$1/\sqrt{\lambda}$ and a mass of order $N_c\lambda$ in units of
$M_{KK}$, the Kaluza-Klein scale~\cite{Sakai3}. The static
Skyrmion is just the instanton holonomy in the $z$-direction, with
a larger size of order $\lambda^0$~\cite{SEITZ}.

In the past the Skyrmion-Skyrmion interaction was mostly
analyzed using the product ansatz~\cite{ANSATZ} or some
variational techniques~\cite{OKA}. While the product
ansatz reveals a pionic tail in the spin and tensor channels,
it lacks the intermediate range attraction in the
scalar channel expected from two-pion exchange. In fact the
scalar potential to order $N_c$ is found to be mostly repulsive,
and therefore unsuited for binding nuclear matter at large $N_c$.
The core part of the Skyrmion-Skyrmion interaction in the
product of two Skyrmions is ansatz dependent.
In~\cite{YAMA} it was shown that the ansatz dependence could
be eliminated in the two-pion range by adding the pion cloud
to the core Skyrmions. In a double expansion using large $N_c$
and the pion-range, a scalar attraction was shown to develop
in the two-pion range in the scalar channel~\cite{YAMA}.
The expansion gets quickly involved while addressing shorter
ranges or core interactions.

In this paper we analyze the two-baryon problem using the D4
two-instanton solution~\cite{BAAL} to order $N_c/\l$.
The ensuing Skyrmion-Skyrmion interaction is essentially
that of the two cores and the meson cloud composed of
(massles) pions and vector mesons. At strong coupling,
holography fixes the core interactions in a way that the
Skyrme model does not. Although in QCD very short ranged
interactions are controlled by asymptotic freedom, the
core interactions at intermediate distances maybe still
in the realm of strong coupling and therefore unamenable
to QCD perturbation theory. In this sense, holography
will be helpful. Also, in holographic QCD the mesonic
cloud including pions and vectors is naturally added
to the core Skyrmions in the framework of semiclassics.
These issues will be quantitatively addressed in this
paper.

In section 2, we review the ADHM construction for one
and two-instanton following on recent work in~\cite{BAAL}.
In section 3, we show
how this construction translates to the one and two
baryon configuration in holography. In section 4, we
construct the bare or core Skyrmion-Skyrmion interaction for
defensive and combed Skyrmions. We unwind
the core Skyrmion-Skyrmion interaction at large separations
in terms of a dominant Coulomb repulsion in regular gauge.
Core issues related to the singular gauge are also discussed.
In section 5, we project the core Skyrmion-Skyrmion contributions
onto the core nucleon-nucleon contributions at large separation
using semiclassics in the adiabatic approximation. In section 6, we
include the effects of the mesonic cloud to order $N_c/\l$ in the
Born-Oppenheimer approximation. At large separations, the cloud
contributions yield a tower of meson exchanges. In section 7,
we summarize the general structure of the NN potential as
a core plus cloud contribution in holographic QCD. Our conclusions
are in section 8. In Appendix A we detail the $k=1,2$ instantons
in the singular gauge.  In Appendix B, we revisit the core
interaction in the singular gauge. In Appendix C, we check our
semiclassical cloud calculations in the regular gauge, using the
strong coupling source theory in the singular gauge. In Appendix D
we detail our nucleon axial-form factor and the extraction of the
axial coupling $g_A$.

\section{YM Instantons from ADHM}

The starting point for baryons in holographic QCD are
instantons in flat $R^3_X\times R_Z$. In this section
we briefly review the ADHM construction~\cite{ADHM} for
SU(2) Yang-Mills instantons. Below SU(2) will be viewed
as a flavor group associated to $\DeDeB$ branes.
For a thorough presentation of the ADHM construction we
refer to~\cite{ADHM1} and references therein.

In the ADHM construction, all the instanton
information is encoded in the matrix-data whose elements
are quaternions $q$. The latters are represented as
\begin{eqnarray}
q \equiv q_M \s^{M} \ ,  \qquad   \s^{M} \equiv (i\tau^{i},\mone) \  ,
\end{eqnarray}
with $M=1,2,3,4$, $\mone \equiv 1_{2\times2} $, and $\t^i$ the
standard Pauli matrices. $q_M$ are four real numbers.
The conjugate ($q^\dg$) and the modulus  ($\| q \|$) of the
quaternion, are defined as
\begin{eqnarray}
&& q^\dg \equiv q_M (\s^M)^\dg  \  ,  \quad  \|q\|^2 \equiv q^{\dg}q =  q q^{\dg} = \abs{\,q} \mone = \sum_M q_M^2 \mone \ ,  \\
&& \mathrm{Re}\, q \equiv \frac{q + q^\dg}{2} = q_0 \s^0 \ , \quad
\mathrm{Im}\, q \equiv \frac{q - q^\dg}{2} = \sum_i q_i \s^i \ ,
\end{eqnarray}
where $\abs{\,q}$ is the determinant of a matrix $q$. For clarity,
our label conventions are: $M,N,P,Q \in \{1,2,3,4 \}$, $\ \m, \n, \rho, \s, \in \{0,1,2,3\}$,
and $\ i,j,k,l \in \{1,2,3\}$ with $z \equiv x_4$. The flavor
$SU(2)$ group indices are $a,b \in \{ 1,2,3 \} $.

The basic block in the matrix-data is the $(1+k) \times k$ matrix, $\D$,
for the charge $k$ instanton
\begin{eqnarray}
\D = \mathbb{A} + \mathbb{B} \otimes x \ ,
\end{eqnarray}
where $\mathbb{A}$ and $\mathbb{B}$ are $x$-independent $(1+k) \times k$ quaternionic matrices
with information on the moduli parameters. We define $x = x_M \s^M$ and
$\mathbb{B} \otimes x$ means that each element $\mathbb{B}$ is multiplied by $x$.
$\mathbb{A}$ and $\mathbb{B}$ are not arbitrary. They follow from the ADHM constraint
\begin{eqnarray}
\D^\dg \D = f^{-1}\otimes \mone \ , \label{finv}
\end{eqnarray}
where $\D^\dg$ is the transpose of the quaternionic conjugate of $\D$. $f$ is
a $k \times k$ invertible quaternionic matrix. $f^{-1}\otimes\mone$
means  each element $f^{-1}$ is multiplied by $\mone$.
The null-space of $\D^\dg$ is 2-dimensional since it has 2 fewer rows than columns.
The basis vectors for this null-space can be assembled into an $(1+k)\times 1$ quaternionic
matrix $U$
\begin{eqnarray}
\D^\dg U = 0 \ ,
\end{eqnarray}
where U is normalized as
\begin{eqnarray}
U^\dg U = \mone \ . \label{Normalize}
\end{eqnarray}

The instanton gauge field $A_\m$ is constructed as
\begin{eqnarray}
A_M = i U^\dg \dell_M U \ ,
\label{AM}
\end{eqnarray}
which yields the field strengths
\begin{eqnarray}
F_{MN} = - 2 \eta_{aMN} U^\dg \mathbb{B} (f \otimes \tau^a ) \mathbb{B}^\dg U \ . \label{FMN}
\end{eqnarray}
Self-duality is explicit from  't Hooft's self-dual eta symbol
\begin{eqnarray}
\eta_{aMN} = -\eta_{aNM} =
\left\{ \begin{array}{ll}
           \e_{aMN}\ & \mathrm{for}\ M, N =1,2,3 \\
           \d_{aM}\ & \mathrm{for}\ N = 4
        \end{array}  \right. \ .
\end{eqnarray}
The action density, $\Tr F_{MN}^2$, can be calculated directly  from $f$,
without recourse to the null-space $U$ and $F_{MN}$~\cite{Osborn}
\begin{eqnarray}
\Tr F_{MN}^2 =  \Box^2 \log \abs{f} \ , \label{OsbornForm}
\end{eqnarray}
where $\Box \equiv \dell_M^2$, $\Box^2 = \dell_N^2 \dell_M^2$, and $\abs{f}$ is the determinant of $f$.

\subsection{$k=1$ instanton}

\indent The $k=1$ instanton in the regular gauge is encoded in a quaternionic matrix $\D$
\begin{eqnarray}
\D \equiv \begin{pmatrix}
       \l  \\
       -x+X
     \end{pmatrix} \ , \qquad
     \D^{\dagger} \equiv \begin{pmatrix}
                 \l^\dg & (-x+X)^\dg
               \end{pmatrix} \ ,
\end{eqnarray}
which yields
\begin{eqnarray}
f^{-1} = \rho^2 + (x_M-X_M)^2 \ , \label{onefinv}
\end{eqnarray}
after using (\ref{finv}). Here $\rho$ ($=\!\!\sqrt{\l_M^2}$) is the size
and $\{X_M\}$ is the position of the one instanton.
The field strength is
\begin{eqnarray}
F_{MN} = W \eta_{aMN}\frac{\t^a}{2}\frac{-4\rho^2}{((x_M-X_M)^2 + \rho^2)^2} W^\dg \ , \label{Fmnone}
\end{eqnarray}
which follows from (\ref{FMN}) with
\begin{eqnarray}
U = \frac{\rho}{\sqrt{(x_M-X_M)^2} + \rho^2} \begin{pmatrix}
                                               -\frac{\l(x-X)^\dg}{\rho^2} \\
                                               \mone \\
                                             \end{pmatrix} W^\dg \ , \qquad
B =\begin{pmatrix}
     0 \\
     -\mone \\
   \end{pmatrix} \ ,
\end{eqnarray}
where $\rho^2 \equiv \l^\dg \l$ and $W \in SU(2)$.
The action density follows from (\ref{Fmnone}) or (\ref{OsbornForm})
\begin{eqnarray}
\Tr F_{MN}^2 = \Box^2 \log f \  = \frac{96\rho^4}{((x_M-X_M)^2+\rho^2)^4}, \label{onedensity}
\end{eqnarray}
which gives the instanton number $\frac{1}{16\pi^2}\int d^4x \Tr F_{MN}^2 = 1$ by
self duality. The $k=1$ instanton in the singular gauge is detailed in Appendix A.

\subsection{$k=2$ instanton}

\indent A charge $2$ ($k=2$) instanton in the regular gauge
is encoded in a quaternionic matrix $\D$~\cite{BAAL}
\begin{eqnarray}
\D \equiv \begin{pmatrix}
       \l_1 & \l_2 \\
       -\big[x-(X+D)\big] & u \\
       u & -\big[x-(X-D)\big] \\
     \end{pmatrix} \ ,
\end{eqnarray}
where the coordinates $x_M$ are defined as $x=x_M \s^M$, and the moduli parameters are
encoded in the free parameters $\l_1, \l_2, X, D$:
$\abs{\l_{i}}  \equiv  \rho_{i} \mone $ are the size parameters,
$\l_1^{\dg}\l_2/(\rho_1\rho_2) \in SU(2)$ is the relative gauge orientation,
and $X\pm D$ is the location of the constituents. $u$ is not a free parameter
and will be determined in terms of other moduli parameters by the
ADHM constraint (\ref{finv}).

Since we are interested in the relative separation we set $X=0$, so that
\begin{eqnarray}
\D = \begin{pmatrix}
       \l_1 & \l_2 \\
       D-x & u \\
       u & -D-x \\
     \end{pmatrix} \ , \qquad
     \D^{\dagger} \equiv \begin{pmatrix}
                 \l_1^\dg & (D-x)^\dg  & u^\dg \\
                  \l_2^\dg & u^\dg & (-D-x)^\dg \\
               \end{pmatrix} \ , \label{Ddagger}
\end{eqnarray}
which yields
\begin{eqnarray}
\D^{\dagger}\D = \begin{pmatrix}
                   \|\l_1\|^2 + \|{x-D}\|^2 + \|{u}\|^2 & \l_1^\dg \l_2 + D^\dg u - u^\dg D - (x^\dg u + u^\dg x) \\
                   \big[\l_1^\dg \l_2 + D^\dg u - u^\dg D - (x^\dg u + u^\dg x)\big]^\dg & \|\l_2\|^2 + \|x+D\|^2 + \|u\|^2 \\
                 \end{pmatrix} \ . \label{DdgD}
\end{eqnarray}
The ADHM constraint (\ref{finv}) implies that each entry must be proportional to $\mone$.
The diagonal terms satisfy the constraint. The off-diagonal entries are also proportional
to $\mone$ provided that $u$ is chosen to be
\begin{eqnarray}
u = \frac{D\Lambda}{2\abs{D}^2} + \g D \ , \qquad \Lambda \equiv  \mathrm{Im} (\l_2^\dg \l_1)
= \half (\l_2^\dg \l_1 - \l_1^\dg \l_2 ) \ , \label{u}
\end{eqnarray}
with $\g$ an arbitrary real constant. The coordinate $u$ is the inverse of the
coordinate $D$. It plays the role of the dual distance.
Throughout we follow~\cite{BAAL} and
choose $\g=0$ for a {\it physical identification of the moduli parameters}.
By that we mean a $k=2$ configuration which is the closest to the superposition
of two instantons in the regular gauge at large separation. In Appendix A, we
briefly discuss a minimal $k=2$ configuration in the singular gauge.

Inserting $u$  into (\ref{DdgD}) yields
\begin{eqnarray}
f^{-1}  = \begin{pmatrix}
                   \rho_1^2 + (x_M  -  D_M)^2 + \frac{\rho_1^2\rho_2^2-
(\l_{1} \cdot \l_{2})^2}{4D_M^2} &  \l_{1} \cdot \l_{2} + 2x \cdot u \\
                   \l_{1} \cdot \l_{2} + 2x \cdot u & \rho_2^2 + (x_M  +  D_M)^2
+ \frac{\rho_1^2\rho_2^2- (\l_{1} \cdot \l_{2})^2}{4D_M^2} \\
                 \end{pmatrix}  \ ,
\end{eqnarray}
where we introduced the notation $ q \cdot p $ for two quaternions $q$ and $p$
\begin{eqnarray}
q \cdot p \equiv \sum_M q_{M} p_{M} \ .
\end{eqnarray}
$\rho_{i} = \sqrt{ \l_{i} \cdot  \l_{i} } $ are the size parameters,
$\pm D_M$ the relative positions of the instantons, and
\begin{eqnarray}
2 x \cdot u = \frac{1}{D\cdot D} \left[ (\l_2 \cdot D)(\l_1 \cdot x)
- (\l_1 \cdot D)(\l_2 \cdot x) - \e^{MNPQ}(\l_2)_M (\l_1)_N D_P x_Q  \right] \ .
\end{eqnarray}
We made use of the identity
\begin{eqnarray}
&& \s^P \bar{\s}^{MN} = \d^{PM}\s^{N} - \d^{PN}\s^{M} - \e^{PMNQ}\s^{Q} \ , \nn \\
&& \bar{\s}^{MN} \equiv \half(\bar{\s}^M \s^N - \bar{\s}^N \s^M) \ , \quad \e^{1234} = 1 \ .
\end{eqnarray}

\subsection{Explicit Parametrization}

Without loss of generality, we may choose the moduli parameters to be
\begin{eqnarray}
\l_1 = \rho_1\left(0,0,0,1 \right)\ , \quad  \l_2 = \rho_2\left(\wh{\theta}_a \sin\!\abs{\,\theta} , \cos\!\abs{\,\theta} \right) \ ,
\quad D = \left( \frac{d}{2},0,0,0 \right)  \ ,  \label{parameter}
\end{eqnarray}
with $a=1,2,3$, $\abs{\,\theta} \equiv \sqrt{(\theta_1)^2
+ (\theta_2)^2 +(\theta_3)^2}$ and $\wh{\theta}_a \equiv \frac{\theta_a}{\abs{\,\theta}} $.
The spatial $x^1$ axis is chosen as the separation axis of two instantons at large distance
$d$. The flavor orientation angles ($\theta_a$) are relative to the $\l_1$ orientation.
We assign an $SU(2)$ matrix $U$  to the relative angle orientations in flavor space
\begin{eqnarray}
U \equiv \frac{\l_1^\dagger \l_2}{\rho_1 \rho_2} = e^{i\theta_a\tau^a} \in SU(2) \ ,
\end{eqnarray}
which is associated with the orthogonal $SO(3)$ rotation matrix $R$ as
\begin{eqnarray}
R_{ab} &=& \half \Tr \left( \tau_a  U \tau_bU^\dagger \right) \nn \\
       &=& \d_{ab}\cos 2\!\abs{\,\theta} + 2\wh{\theta}_a \wh{\theta}_b \sin^2\!\abs{\,\theta}
+ \e_{abc} \wh{\theta}_c \sin2\!\abs{\,\theta}  \ . \label{SO3}
\end{eqnarray}
For instance $R_{ab}$ reads
\begin{eqnarray}
\begin{pmatrix}
  \cos 2\theta_3 & \sin 2\theta_3 & 0 \\
  -\sin 2\theta_3 & \cos 2\theta_3& 0 \\
  0 & 0 & 1 \\
\end{pmatrix} \ , \qquad
\begin{pmatrix}
  1 & 0 & 0 \\
  0 & \cos 2\theta_1 & \sin 2\theta_1  \\
  0 & -\sin 2\theta_1 & \cos 2\theta_1  \\
\end{pmatrix} \ ,
\end{eqnarray}
for $\theta_1 = \theta_2 = 0$ and $\theta_2 = \theta_3 = 0$ respectively.
Note the double covering in going from SU(2) to SO(3).

In this coordination for the moduli space,
\begin{eqnarray}
&& \Lambda = \rho_1 \mathrm{Im}(\l_2^\dg) = \rho_1 \rho_2
( -i \wh{\theta}_a \tau^a \sin\!\abs{\,\theta}) \ , \label{parameter1} \\
&& u = \frac{D\Lambda}{2\abs{D}^2} = \frac{i\t^1 \Lambda}{d}
= \frac{\rho_1 \rho_2}{d} \sin\!\abs{\,\theta} \wh{\theta}_a \t^1 \t^a \ , \\
&& u_M =  \frac{\rho_1 \rho_2}{d} \sin\!\abs{\,\theta} \left(0, - \wh{\theta}_3 , \wh{\theta}_2, \wh{\theta}_1 \right) \ , \\
&&  x \cdot u = \frac{\rho_1\rho_2 \sin\abs{\,\theta}}{d} \left(\wh{\theta}_1 x_4 +\wh{\theta}_2 x_3 - \wh{\theta}_3 x_2 \right)  \ ,
\end{eqnarray}
and the inverse potential $f^{-1}$ is written as
\begin{eqnarray}
&& f^{-1} = \begin{pmatrix}
  g_- + A & B \\
  B & g_+  + A \\
   \end{pmatrix}  \label{finv2} \ , \\
&& \quad g_{\pm} \equiv x_\a^2   + \left(x_1 \pm  \frac{d}{2}\right)^2  + \rho^2 \ , \quad x_\a^2 \equiv  x_2^2 + x_3^2 + x_4^2 \ , \\
&& \quad  A \equiv \frac{\rho^4 \sin^2\abs{\,\theta}}{d^2} \ , \quad B \equiv  \rho^2 \left(\cos\abs{\,\theta} + \frac{2}{d} \sin\abs{\,\theta}
  \left[\wh{\theta}_1 z +\wh{\theta}_2 x_3 - \wh{\theta}_3 x_2  \right] \right) \ ,
\end{eqnarray}
with $\rho \equiv \rho_1 = \rho_2$. The action density can be assessed using (\ref{OsbornForm}).
In terms of this notation, for the $k=1$ instanton in the regular gauge (\ref{onefinv}), the logarithmic potential $\log \abs{f}$ is
\begin{eqnarray}
 \log f_{\pm} =  -  \log g_{\pm} \ ,  \label{fpm1}
\end{eqnarray}
where the subscript $\pm$ refers to the position $\mp \frac d 2$ of the instanton
along the $x_1$ axis. For the $k=2$ instanton (\ref{finv2}), we have
\begin{eqnarray}
 \log \abs{f_{-+}} &\equiv&  -  \log \left[ \left( g_- +
 A \right) \left( g_+  + A  \right)  - B^2 \right] \ .  \label{fpm2}
\end{eqnarray}

\subsection{Asymptotics }

To understand in details the structure of the $k=2$ instanton it is best
to work out its asymptotic form for the case $d/\rho\gg 1$. For that we
use (\ref{FMN}) in the special case
\begin{eqnarray}
&& F_{iz} = - 2  U^\dg \mathbb{B} (f \otimes \tau^i ) \mathbb{B}^\dg U \ , \label{Fiz}
\end{eqnarray}
with
\begin{eqnarray}
\mathbb{B} = \begin{pmatrix}
      0 & 0 \\
      -\mone & 0 \\
      0 & -\mone \\
    \end{pmatrix} \label{B} \ .
\end{eqnarray}
Below, we will show that the field strength $F_{iz}$ sources the
pion-nucleon coupling in the axial gauge $A_z=0$ for the quantum
fluctuations. The asymptotics is useful for a physical identification
of the coset parameters.

Near the singularity center with $x=D$, (\ref{Ddagger}) approximates to
\begin{eqnarray}
\D^{\dagger} \approx \begin{pmatrix}
             \l_1^\dg & 0  & u^\dg \\
             \l_2^\dg & u^\dg & -2D^\dg \\
             \end{pmatrix} \ ,
\end{eqnarray}
whose null vector $U$ is
\begin{eqnarray}
U \approx \begin{pmatrix}
      -\frac{1}{\rho_1}u^\dg \\
      \inv{\abs{u}^2}\inv{\rho_1} u\left(\l_2^\dg u^\dg + 2\rho_1 D^\dg   \right)  \\
      \mone \\
    \end{pmatrix}   D\Lambda^\dg D^\dg .
\end{eqnarray}
From (\ref{Normalize}) and (\ref{u}) it follows that
%
%
\begin{eqnarray}
U\approx\begin{pmatrix}
        0 \\
        \mone - \left(\frac{\rho}{d}\right)^2 \frac{1}{2} \sin2\!\abs{\theta} \wh{\theta}_a (i\t^1\t^a\t^1) \\
        \left(\frac{\rho}{d}\right)^2  \sin\abs{\theta} \wh{\theta}_a (i\t^1\t^a\t^1)
      \end{pmatrix} +
        \begin{pmatrix}
          \calo\left(\frac{\rho}{d}\right)^3 \\
          \calo\left(\frac{\rho}{d}\right)^4 \\
          \calo\left(\frac{\rho}{d}\right)^4 \\
        \end{pmatrix} \ . \label{Umatrix}
\end{eqnarray}
We have used the explicit parametrization (\ref{parameter}) and (\ref{parameter1}).
We may expand $f$ near the center $X=D$,
\begin{eqnarray}
f|_{X \approx D}  = \begin{pmatrix}
      \inv{g_+} + \calo\left(\frac{1}{d}\right)^4 & -\frac{\l_1\l_2 + 2x\cdot u}{g_-g_+}+ \calo\left(\frac{1}{d}\right)^4 \\
      -\frac{\l_1\l_2 + 2x\cdot u}{g_-g_+} + \calo\left(\frac{1}{d}\right)^4 & \inv{g_-} + \calo\left(\frac{1}{d}\right)^2 \\
    \end{pmatrix} \ .
\end{eqnarray}
For $X=D$, the leading contributions to $f_{11}, f_{12},$ and $f_{21}$ are of order
$1/d^2$ while that of $f_{22}$ is of order $d^{0}$.

From (\ref{B}) and (\ref{Umatrix}) we have
\begin{eqnarray}
&& U^\dg B|_{X \approx D} = \left(-\mone + \calo\left(\frac{\rho}{d}\right)^2   , \calo\left(\frac{\rho}{d}\right)^2\right)
\equiv \left( \mathfrak{b}_1^\dg , \mathfrak{b}_2^\dg \right) \ ,
\end{eqnarray}
which yields (\ref{Fiz})
\begin{eqnarray}
 F_{iz}|_{X \approx D} &=& -2 U^\dg B (f \otimes \tau^i ) B^\dg U|_{X \approx D} \nn \\
  &=& -2\left( f_{11}\mathfrak{b}_1^\dg\t^i\mathfrak{b}_1  + f_{12}\mathfrak{b}_1^\dg\t^i\mathfrak{b}_2
     + f_{21}\mathfrak{b}_2^\dg\t^i\mathfrak{b}_1  + f_{22}\mathfrak{b}_2^\dg\t^i\mathfrak{b}_2 \right) \nn \\
  &=&  -2 \frac{\t^i}{g_+} + \calo(d^{-4}) \ .  \label{Fiza}
\end{eqnarray}
Thus
\begin{eqnarray}
 F_{iz}^a|_{X \approx D} \approx  -2 \d^{ia} \frac{1}{g_+}
\end{eqnarray}
A rerun of the argument for the center $x = - D$ yields
\begin{eqnarray}
 F_{iz}^a|_{X \approx - D} \approx  -2 R^{ia} \frac{\t^i}{g_-}  \ , \label{Fizb}
\end{eqnarray}
since $U^\dg \sim (0,0,\mone)$. For asymptotic distances $d/\rho\gg 1$ the $k=2$
configuration splits into two independent $k=1$ configurations with relative flavor
orientation $R^{ab}$. This separation makes explicit the physical interpretation of
the coset parameters: $\rho$ the instanton size, $d$ the instanton relative separation,
$u$ the inverse or dual separation and $R$ their relative orientations asymptotically.

\section{Baryons in hQCD}

Baryons in hQCD are sourced by instantons in bulk. The
induced action by pertinent brane embeddings and its
instanton content was discussed in~\cite{Sakai3}.
The 5D {\it effective} Yang-Mills
action is the leading terms in the $1/\l$ expansion of the DBI
action of the D8 branes after integrating out the $S^4$. The 5D
Chern-Simons action is obtained from the Chern-Simons action of
the D8 branes by integrating $F_4$ RR flux over the $S^4$, which
is nothing but $N_C$. The action reads~\cite{Sakai1,Sakai3}
\begin{eqnarray}
&&S = S_{YM} + S_{CS}\ ,  \label{YM-CS}\\
&&S_{YM} = - \k \int d^4x dz \ \Tr \left[\half K^{-1/3}
\calf_{\m\n}^2 + \Mkk^2 K
\calf_{\m z}^2 \right] \ , \label{YM} \\
&&S_{CS} = \frac{N_c}{24\pi^2}\int_{M^4 \times R}
\w_5^{U(N_f)}(\cala) \ , \label{CS}
\label{REDUCED}
\end{eqnarray}
where $\m,\n = 0,1,2,3$ are 4D indices and the fifth(internal)
coordinate $z$ is dimensionless.  There are three things which are
inherited by the holographic dual gravity theory: $\Mkk, \k,$ and
$K$. $\Mkk$ is the Kaluza-Klein scale and we will set $\Mkk = 1$
as our unit. $\k$ and $K$ are defined as
\begin{eqnarray}
\k = {\l N_c} \inv {216 \pi^3} \equiv \l N_c a  \ , \qquad K = 1
+ z^2 \ .
\end{eqnarray}
$\cala$ is the 5D $U(N_f)$ 1-form gauge field and $\calf_{\m\n}$
and $\calf_{\m z} $ are the components of the 2-form field
strength $\calf = \ud \cala -i \cala \wedge \cala$.
$\w_5^{U(N_f)}(\cala)$ is the Chern-Simons 5-form for the $U(N_f)$
gauge field
\begin{eqnarray}
  \w_5^{U(N_f)}(\cala) = \Tr \left( \cala \calf^2
   + \frac{i}{2} \cala^3 \calf - \inv{10} \cala^5
  \right)\ ,
\end{eqnarray}

The exact instanton solutions in warped $x^M$ space are not known.
Some generic properties of these solutions can be inferred from
large $\l$ whatever the curvature. Indeed, since
$\k \sim \l$, the instanton solution with unit topological charge
that solves the full equations of motion, follows from the
YM part only in leading order. It has zero size at infinite $\l$.
At finite $\l$ the instanton size is of order $1/\sqrt{\l}$. The
reason is that while the CS contribution of order $\l^0$ is
repulsive and wants the instanton to inflate, the warping in the
$z$-direction of order $\l^0$ is attractive and wants the instanton
to deflate in the $z$-direction~\cite{RHO,Sakai3}.

These observations suggest to use the flat space instanton
configurations to leading order in $N_c\l$, with $1/\l$ corrections
sought in perturbation theory. The latter is best achieved by
rescaling the coordinates and the instanton fields as
\begin{eqnarray}
&&x^M = \l^{-1/2}\widetilde{x}^M \ , \quad x^0 = \widetilde{x}^0 \ , \nn \\
&&\cala_M = \l^{1/2} \widetilde{\cala}_M \ , \quad \cala_0 = \widetilde{\cala}_0 \ , \nn \\
&& \calf_{MN} = \l \widetilde{\calf}_{MN} \ , \quad \calf_{0M} =
\l^{1/2} \widetilde{\calf}_{0M} \ . \label{Rescaling}
\label{Rescale}
\end{eqnarray}
The corresponding energy density associated to the action (\ref{REDUCED}) reads~\cite{Sakai3}
\begin{eqnarray}
&& E = 8\pi^2\k \left[\inv{16\pi^2} \int d^3\wt{x} d\wt{z} \Tr \wt{F}_{MN}^2\right]  \nn  \\
&& \qquad \ \ +\ \frac{\k}{\l} \int d^3\wt{x} d\wt{z} \left[-\frac{\wt{z}^2}{6}\Tr \wt{F}_{ij}^2
+ \wt{z}^2\Tr \wt{F}_{iz}^2  -\half (\widetilde{\dell}_M \wh{\widetilde{A}}_0)^2 - \inv{32\pi^2a} \wh{\widetilde{A}}_0
\Tr \widetilde{F}_{MN}^2  \right] \ . \label{Energy}
\end{eqnarray}
All quantities are dimensionless in units of $\Mkk$.
The U(1) contribution $\wh{\widetilde{A}}_0$ follows from the equation of motion~\cite{Sakai3}
\begin{eqnarray}
\wt{\Box} \wh{\widetilde{A}}_0 = \frac{1}{32\pi^2 a} \Tr \widetilde{F}_{MN}^2  \ .
\end{eqnarray}
The $\wh{\widetilde{A}}_0$ field can be obtained in closed form using (\ref{OsbornForm}),
\begin{eqnarray}
\wh{\widetilde{A}}_0 = \frac{1}{32\pi^2 a} \wt{\Box} \log \abs{f} \ .
\end{eqnarray}
According to (\ref{Rescaling}) {\it both} the size of
the instanton $\rho$ and the distance $d$ between two instantons
are rescaled, i.e. $\wt{\rho} = \sqrt{\l}\rho$
and $\wt{d} = \sqrt{\l}d$. While the size $\wt{\rho}$ is fixed
to $\wt{\rho_0}$ (see below) by the energy minimization process, the
distance is not. Therefore, when discussing the energy at the
subleading order, the distance $\wt{d}$ is always short for
$\sqrt{\l} d$. It will be recalled whenever appropriate.
The first term in (\ref{Energy}) is $8\pi^2\k$ $\times$ instanton number, which is identified with the bare soliton
mass. The second term ($\equiv \D E$) is subleading and corresponds to the correction to the mass or the interaction
energy
\begin{eqnarray}
 \D E &=& \frac{\k}{\l} \int d^3\wt{x} d\wt{z} \left[-\frac{\wt{z}^2}{6}\Tr \wt{F}_{ij}^2
+ \wt{z}^2\Tr \wt{F}_{iz}^2  -\half (\wt{\dell}_M\wh{\wt{A}}_0)^2 - \inv{32\pi^2a} \wh{\wt{A}}_0 \Tr \wt{F}_{MN}^2  \right] \nn \\
 &=& \frac{\k}{6\l} \int d^3\wt{x} d\wt{z} \left(\wt{z}^2-\frac{3^7
\pi^2}{2^4} \wt{\Box} \log \abs{f} \right) \wt{\Box}^2 \log \abs{f} \ , \label{IntEnergy}
\end{eqnarray}
where we used the self-duality, $\Tr \wt{F}_{ij}^2 = 2 \Tr \wt{F}_{iz}^2 = \half \Tr \wt{F}_{MN}^2$,
and integrated  $(\dell_M\wh{\wt{A}}_0)^2$ by part so that it can be reduced to the form $\wh{\wt{A}}_0 \Tr \wt{F}_{MN}^2 $.

\newpage

\subsection{One baryon}

\begin{figure}[]
  \begin{center}
  \includegraphics[width=7cm]{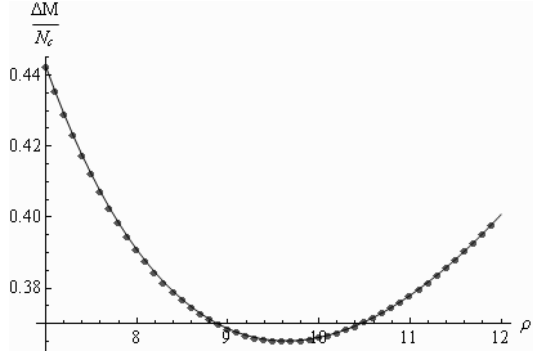}
  \caption{$\Delta M/N_c$: solid (exact) and dotted (numerical).}
  \label{Fig:mass}
  \end{center}
\end{figure}

\indent The one baryon solution is the $k=1$ instanton.
From (\ref{onefinv}) it follows that
\begin{eqnarray}
f^{-1} = \wt{\rho}^2 + \wt{x}_M^2 \ , \label{finv1}
\end{eqnarray}
for $k=1$. We have set $\wt{X}_i = 0$ by translational symmetry.
We have also set $\wt{X}_4=0$ as a finite $\wt{X}_4$ cots energy~\cite{Sakai3}. Thus
\begin{eqnarray}
  && \wt{\Box} \log f = -4 \frac{\wt{x}_M^2 + 2\wt{\rho}^2}{(\wt{x}_M^2+\wt{\rho}^2)^2} \ , \\
  && \wt{\Box}^2 \log f =  \frac{96\wt{\rho}^4}{(\wt{x}_M^2+\wt{\rho}^2)^4} \ .
\end{eqnarray}
The mass correction $\D M \equiv \D E$, reads
\begin{eqnarray}
 \D M(\rho) &=& \frac{\k}{6\l} \int d^3\wt{x} d\wt{z} \left( \wt{z}^2+\frac{3^7
\pi^2}{4}\frac{\wt{x}_M^2 + 2\wt{\rho}^2}{(\wt{x}_M^2+\wt{\rho}^2)^2} \right)
  \frac{96\wt{\rho}^4}{(\wt{x}_M^2+\wt{\rho}^2)^4}  \label{MassCorrection0} \\
  &=&  \frac{8\pi^2\k}{\l} \left(\frac{\wt{\rho}^2}{6} + \frac{1}{320\pi^4 a^2}\inv{\wt{\rho}^2}    \right) \ . \label{MassCorrection}
\end{eqnarray}
It depends on the size $\wt{\rho}$ as plotted in Fig.~\ref{Fig:mass}.
All integrals in $\D M$ are analytical, since
$\wt{\Box} \log \abs{f}$ and $\wt{\Box}^2 \log \abs{f}$ are simple. For
$k=2$ the expressions for $\D M$ are more involved and require
numerical unwinding. As a prelude to these numerics, we have
carried the integrals in (\ref{MassCorrection0}) both analytically
and numerically as illustrated in~Fig.\ref{Fig:mass}.

The one-instanton stabilizes for
\begin{eqnarray}
\wt{\rho}_0 = \sqrt{\frac{1}{8\pi^2a} \sqrt{\frac{6}{5}}} \sim 9.64 \ ,
\label{SIZE}
\end{eqnarray}
with a mass correction
\begin{eqnarray}
\D M(\wt{\rho}_0 \sim 9.64) \sim 0.365 \ .
\end{eqnarray}
We recall that the physical instanton size $\rho_0=\wt{\rho}_0/\sqrt{\l}$ following
the unscaling as detailed above.

\subsection{Two baryon}

\indent The two baryon solution corresponds to the $k=2$ instanton.
The corresponding potential $f$ for the $k=2$ instanton is given
in (\ref{finv2}) and yields
\begin{eqnarray}
\Tr \wt{F}_{\m\n}^2 &=&  \wt{\Box}^2 \log \abs{f} \nn \\
  &=&  - \wt{\Box}^2  \log \left[ \left(g_-(\wt{x}_M) + \frac{\wt{\rho}_1^2\wt{\rho}_2^2 \sin^2\abs{\,\theta}}{\wt{d}^2}\right)
\left( g_+(\wt{x}_M)
+ \frac{\wt{\rho}_1^2\wt{\rho}_2^2 \sin^2\abs{\,\theta}}{\wt{d}^2}  \right) \right. \nn \\
  && \qquad \qquad  \left. - \wt{\rho}_1^2 \wt{\rho}_2^2 \left(\cos\abs{\,\theta}
+ \frac{2}{\wt{d}} \sin\abs{\,\theta}  \left[\wh{\theta}_1 \wt{x}_0
+\wh{\theta}_2 \wt{x}_3 - \wh{\theta}_3 \wt{x}_2  \right] \right)^2 \right]  \ . \label{ActionDensityGeneral}
\end{eqnarray}
Its leading contribution in (\ref{Energy}) is
\begin{eqnarray}
8\pi^2\k \left[\inv{16\pi^2} \int d^3\wt{x} d\wt{z} \Tr \wt{F}_{MN}^2\right]  = 2 \times 8\pi^2\k \nn \ ,
\end{eqnarray}
as expected by self-duality. To order $N_c\l$ the energy of the 2-baryon
system is just $2M_0$ or twice the bare soliton mass. There is complete
degeneracy in the moduli parameters $\wt{d}$ and $\theta_a$. This degeneracy
is lifted at order $N_c\l^0$, which is the next contribution in (\ref{Energy}).
This will be detailed below.

\begin{figure}[]
  \begin{center}
  \includegraphics[width=9cm]{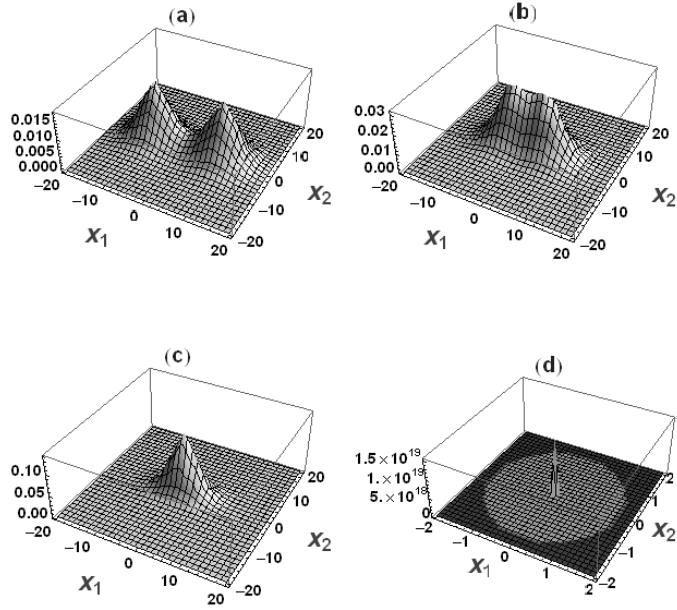}
  \caption{Defensive Skyrmions: (a) $\wt{d}=2$, (b) $\wt{d}=\sqrt{2}$, (c) $\wt{d}=1$, (d) $\wt{d}=10^{-5}$}
  \label{Fig:par}
  \end{center}
\end{figure}

For two parallel instantons $\abs{\,\theta}=0$
and the instanton action density (\ref{ActionDensityGeneral}) reads
\begin{eqnarray}
\Tr \wt{F}_{\m\n}^2  &=&  - \wt{\Box}^2 \log \left[   g_-(\wt{x}_M) g_+(\wt{x}_M) - \wt{\rho}_1^2\wt{\rho}_2^2  \right] \ .
\label{ActionDensityPara}
\end{eqnarray}
The baryon number distribution in space follows from
\begin{eqnarray}
B({x})=\frac 1{16\pi^2}\int_{-\infty}^{+\infty} \,d{z}\,\Tr {F}_{\m\n}^2 \ ,
\label{BARYON}
\end{eqnarray}
which integrates to 2. Since the instanton in bulk is localized
near $z\approx \rho\approx 1/\sqrt{\l}$, we may approximate the
integral by the value of the integrand for $z\approx 0$, or
$B(x)\approx \Tr \wt{F}_{\m\n}^2 (z\approx 0)/16\pi^2$.
In  Fig.~\ref{Fig:par} we show $\Tr \wt{F}_{\m\n}^2$ for $\wt{z}=\wt{x}_3=0$
and $\wt{\rho}_1=\wt{\rho}_2=9.64$ for various separations $\wt{d}$ in the
$(\wt{x}_1,\wt{x}_2)$ space for two paralell Skyrmions. The separation
is in units of the size $\wt{\rho}_0=9.64$. For small separations a narrow
Skyrmion develops on top of the broad Skyrmion. The configuration
is maximally repulsive (defensive Skyrmions).

\begin{figure}[]
  \begin{center}
  \includegraphics[width=9cm]{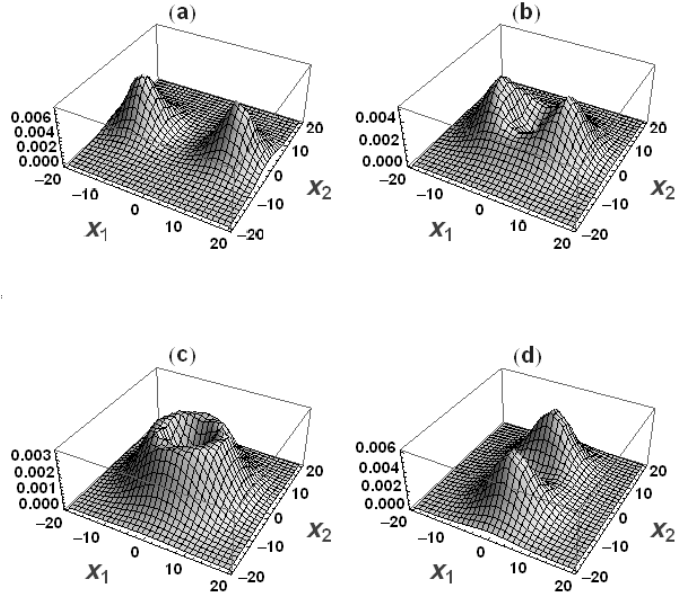}
  \caption{Combed Skyrmions: (a) $\wt{d}=2.5$, (b) $\wt{d}=1.7$, (c) $\wt{d}=\sqrt{2}$, (d) $\wt{d}=1$  }
  \label{Fig:per}
  \end{center}
\end{figure}

A paralell and antiparalell Skyrmion (combed Skyrmions) corresponds
to the choice $\theta_1=\theta_2=0$ and  $\theta_3=\frac{\pi}{2}$
or $\abs{\,\theta} = {\pi}/{2}$. This is a $\pi$ rotation
along $x_3$ in the SO(3) notation (\ref{SO3}). The resulting
instanton action density (\ref{ActionDensityGeneral}) reads
\begin{eqnarray}
\Tr \wt{F}_{\m\n}^2
  &=&  - \wt{\Box}^2  \log \left[ \left(   g_-(\wt{x}_M)   + \frac{\wt{\rho}_1^2\wt{\rho}_2^2}{\wt{d}^2}\right)
   \left(  g_+(\wt{x}_M)   + \frac{\wt{\rho}_1^2\wt{\rho}_2^2}{\wt{d}^2}  \right)
    - 4 \frac{\wt{\rho}_1^2\wt{\rho}_2^2}{\wt{d}^2}  \wt{x}_2^2  \right]\ . \label{ActionDensityPerp}
\end{eqnarray}
In Fig.~\ref{Fig:per} we show the baryon density in the plane $(x_1,x_2)$ for various
separations in units of the instanton size with $\wt{\rho}_1 = \wt{\rho}_2 = 9.64$.
For large separation two lumps form along the $x^1$ axis.  For smaller separation
the two lumps are seen to form in the orthogonal or $x_2$ direction. In between a hollow
baryon 2 configuration is seen which is the precursor of the donut seen in the baryon
number 2 sector of the Skyrme model~\cite{DONUT}. The concept of $\wt{d}$ as a separation
at small separations is no longer physical given the separation taking place in the
transverse direction. What is physical is the dual distance $u$ in the transverse plane.

\begin{figure}[]
  \begin{center}
  \includegraphics[width=9cm]{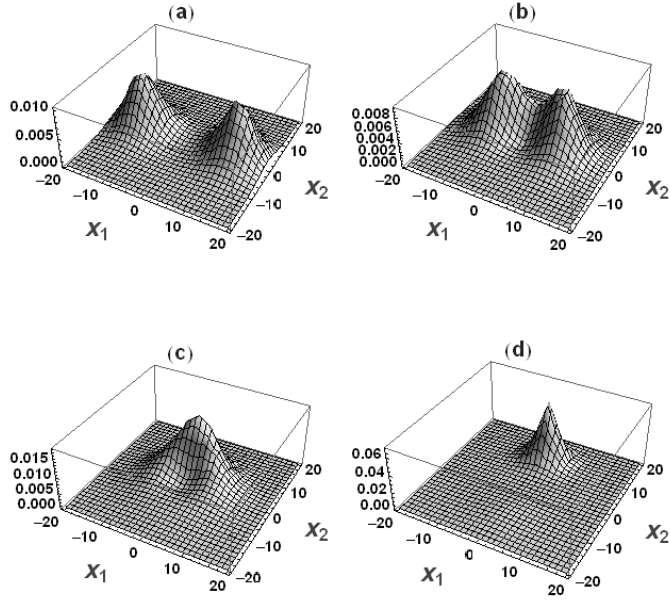}
  \caption{Orthogonal Skyrmions: (a) $\wt{d}=2.5$, (b) $\wt{d}=1.7$, (c) $\wt{d}=1.2$, (d) $\wt{d}=0.6$.}
  \label{Fig:45}
  \end{center}
\end{figure}

For two Skyrmions orthogonal to each other, the choice of angles is
$\theta_1 = \theta_2 = 0, \theta_3 = \frac{\pi}{4}$. The corresponding
action density is given by (\ref{ActionDensityGeneral})
\begin{eqnarray}
\Tr \wt{F}_{\m\n}^2
 &=&  - \wt{\Box}^2  \log \left[ \left(  g_-(\wt{x}_M)    + \frac{\wt{\rho}_1^2\wt{\rho}_2^2 \sin^2\theta_3}{\wt{d}^2}\right)
 \left(  g_+(\wt{x}_M)  + \wt{\rho}_2^2  + \frac{\wt{\rho}_1^2\wt{\rho}_2^2 \sin^2\theta_3}{\wt{d}^2}  \right) \right. \nn \\
  && \qquad \qquad  \left. - \wt{\rho}_1^2 \wt{\rho}_2^2 \left(\cos\theta_3
- \frac{2\wt{x}_2}{\wt{d}} \sin\theta_3  \right)^2 \right]  \ , \label{theta3}
\end{eqnarray}
which is also seen to reduce to (\ref{ActionDensityPara}) and (\ref{ActionDensityPerp})
for $\theta_3 = 0\ \mathrm{or}\ \pi $ and $\theta_3 = \pi/2$ respectively.
The $\theta_3 = \frac{\pi}{4} $ is our two orthogonal Skyrmions. This configuration
is shown in~Fig.(\ref{Fig:45}). For small separations a narrow Skyrmion  develops
on top of a broad one, a situation reminiscent of the Defensive Skyrmion configuration
above. This situation can be seen in many other relative orientations and is somehow
generic.

\section{Skyrmion-Skyrmion Interaction}

The Skyrmion-Skyrmion interaction in hQCD is of order $N_c/\l$ and it
follows from  (\ref{IntEnergy}) which is the second term in (\ref{Energy}).
The baryon two minimum energy configuration should follow by minimizing this contribution in
the 6-dimensional coset space $\rho, d, \theta$. This will be reported
elsewhere. Instead, we report on the interaction energy between
two Skyrmions versus their separation for a size fixed in the baryon
1 sector and different relative orientations $\theta_a$. In the
adiabatic quantization scheme, $\theta_a$ are raised to collective
coordinates. They are not fixed by minimization. This approach will
be subsumed here. We note that the mass shift are of order $N_c \l^0$.

\subsection{General}

\begin{figure}[]
  \begin{center}
  \includegraphics[width=10cm]{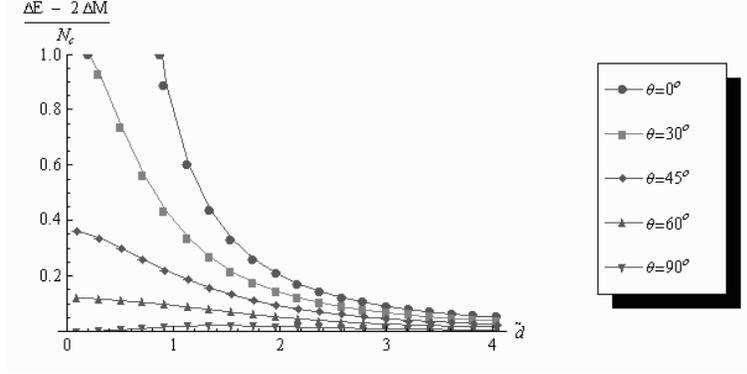}
  \caption{Skyrmion-Skyrmion interaction in regular gauge.}
  \label{Fig:potential}
  \end{center}
\end{figure}

\begin{figure}[]
  \begin{center}
  \includegraphics[width=6cm]{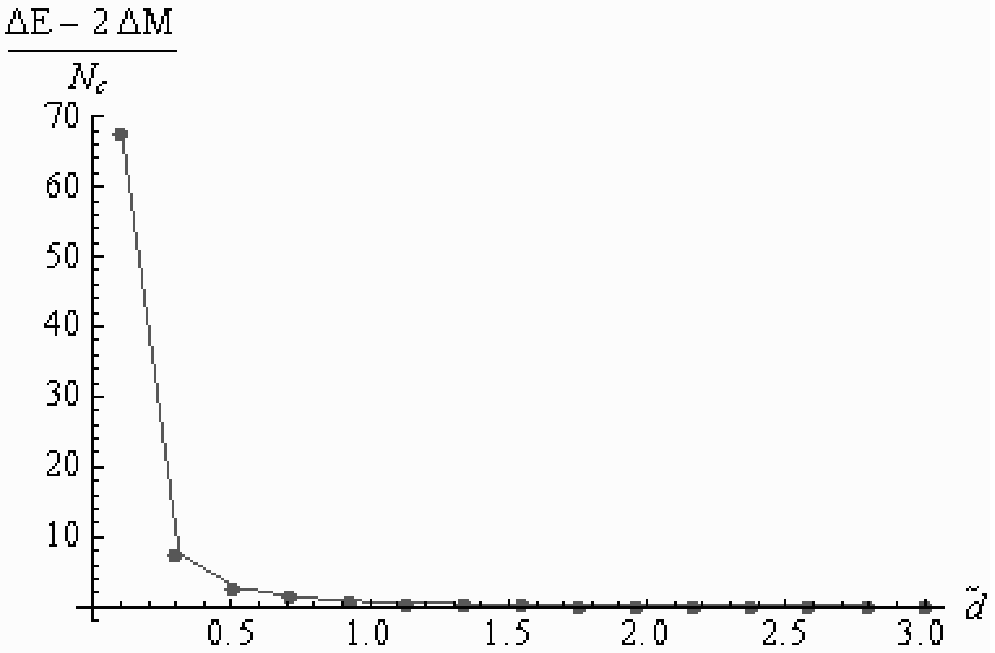}
  \includegraphics[width=6cm]{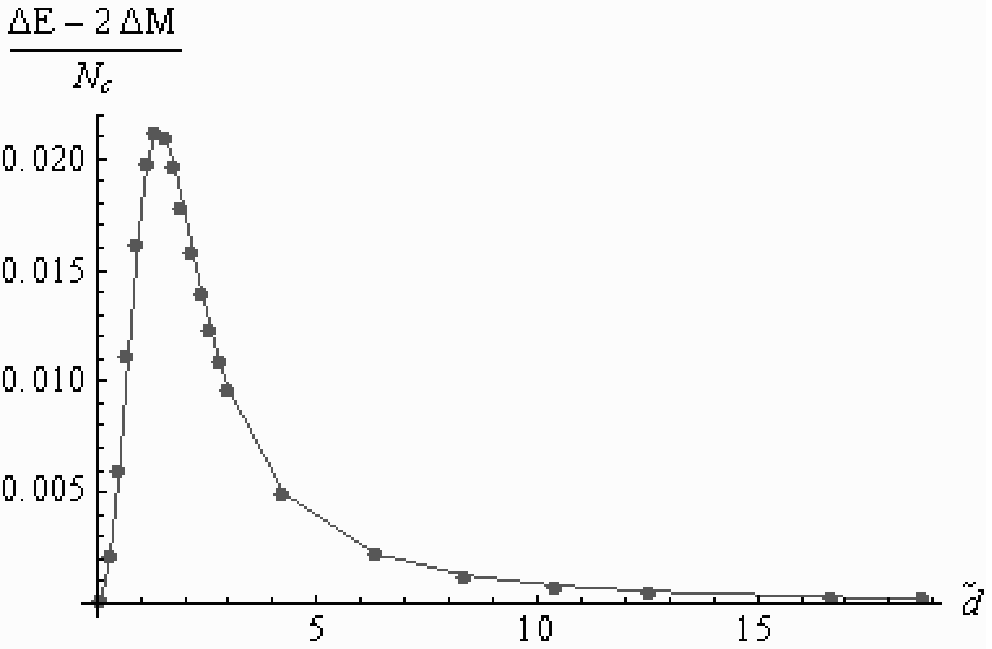}
  \caption{Skyrmion-Skyrmion interaction: Defensive (left) and Combed (right)}
  \label{Fig:potential1}
  \end{center}
\end{figure}

Consider the case where $\theta_1=\theta_2=0$ and $\theta_3 \ne 0 $,
with fixed sizes $\wt{\rho}_1 = \wt{\rho}_2 = \wt{\rho}_0$. Here $\wt{\rho}_0$ is the
value fixed by minimization in the 1 Skyrmion sector (\ref{SIZE}).
In Fig.~(\ref{Fig:potential}) we show the interaction energy
$(\Delta E-2\Delta M)/N_c$ versus the relative distance $d$ in
units of the instanton size, where
\begin{eqnarray}
 && \D E  = \frac{\k}{6\l} \int d^3\wt{x} d\wt{z} \left(\wt{z}^2-\frac{3^7  \pi^2}{2^4}
 \wt{\Box} \log \abs{f} \right) \wt{\Box}^2 \log \abs{f}  \ , \label{DE} \\
 && \abs{f} =   \left(  g_-(\wt{x}_M)  + \frac{\wt{\rho}_1^2\wt{\rho}_2^2 \sin^2\theta_3}{
 \wt{d}^2}\right)
 \left(  g_+(\wt{x}_M) + \frac{\wt{\rho}_1^2\wt{\rho}_2^2 \sin^2\theta_3}{\wt{d}^2}  \right)  \nn \\
  && \qquad  \quad  - \wt{\rho}_1^2 \wt{\rho}_2^2 \left(\cos\theta_3
- \frac{2\wt{x}_2}{\wt{d}} \sin\theta_3  \right)^2 \ .
\end{eqnarray}
The interaction energy is repulsive
for all values of $\theta_3$. The repulsion decreases for $\theta_3$
in the range $0\rightarrow \pi/2$, that is from the defensive to
combed configuration. The defensive or $\theta_3=\pi/2$ is still
repulsive even for small relative distances, as the two Skyrmions
separate in the transverse direction. In Fig.~(\ref{Fig:potential1})
we show separatly the interaction energy for the defensive
configuration (left) and combed configuration (right). The
repulsion is seen to drop by 3 orders of magnitude.

The core interaction is modified in the singular gauge as we detail
in Appendix A and B. In Fig.~(\ref{Fig:potentials}) we show the analogue
of Fig.~(\ref{Fig:potential}) in the singular gauge. The switch from
repulsion to attraction follows from the switch from repulsive Coulomb
(regular gauge) to attractive dipole (singular gauge) interactions. The
plot is versus $\tilde{d}$ which is the rescaled distance in units of
the rescaled size $\tilde{\rho}$. In the unscaled distance $d$, the dipole
attraction is of order $N_c/\l^4$ and subleading.

\begin{figure}[]
  \begin{center}
  \includegraphics[width=10cm]{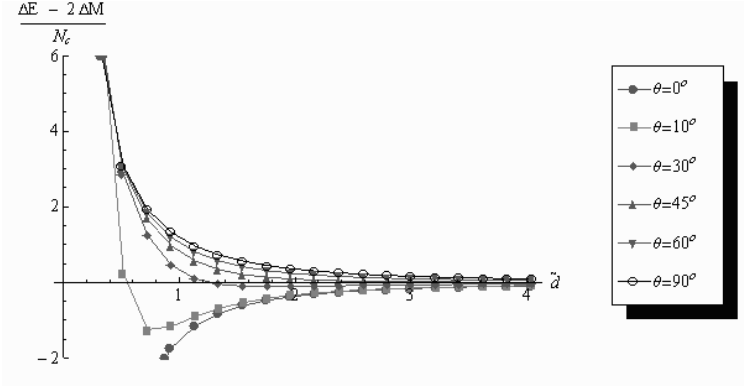}
  \caption{Skyrmion-Skyrmion interaction in singular gauge   }
  \label{Fig:potentials}
  \end{center}
\end{figure}

\subsection{Interaction at Large Separation}

To understand the nature of the Skyrmion-Skyrmion interaction to order
$N_c/\l$ as given by the classical instantons in bulk, we now
detail it for large separations between the instanton cores, i.e.
$d\gg \rho$ but still smaller than the pion range (which is infinite
for massless pions). We recall that the interaction follows from
the subleading contribution in~(\ref{Energy}), which can be split
\begin{eqnarray}
  \D E[f]  &=& N_c b \left(C[f] + c\, D[f] \right)   \ ,   \label{Energy1} \\
     C[f] &\equiv&  \int d^3\wt{x} d\wt{z}\, \wt{z}^2\, \wt{\Box}^2 \log \abs{f}   \ ,\\
     D[f] &\equiv&  -\int d^3\wt{x} d\wt{z}\, (\,\wt{\Box}^2 \log \abs{f}\,) \, \inv{\wt{\Box}} \, (\, \wt{\Box}^2 \log \abs{f} \,)    \ ,
\end{eqnarray}
with $b=\frac{1}{6\cdot 216 \pi^3}$ and $c \equiv \frac{3^7  \pi^2}{2^4}$.

For large separations between the cores or $d \gg \rho$, we have from (\ref{fpm2})
\begin{eqnarray}
 \log \abs{f_{-+}} &=&  -  \log \left[  (g_- g_+) \left( 1 +  A \frac{ g_-  +  g_+ }{g_- g_+}
+ \frac{A^2 - B^2}{g_- g_+}\right) \right]  \nn \\
                   &\approx&  - \log g_- -\log g_+ -  A \frac{ g_-  +  g_+ }{g_- g_+} + \frac{  B^2}{g_- g_+} \ ,
\end{eqnarray}
after dropping the $A^2$ contribution as it is subleading in $\rho/d$.
We note that after fixing the size of the single instanton to $\wt{\rho}_0$
and {\it unscaling} the distance $\wt{d}$ as we indicated above, the
expansion $\wt{\rho}/\wt{d}$ is an expansion in $\wt{\rho}_0/(\sqrt{\l}d)$.

The Skyrmion-Skyrmion core interaction follows from
\begin{eqnarray}
V = \D E[f_{-+}] - \D E[f_-] - \D E[f_+] \ ,
\end{eqnarray}
after subtraction of the classical self-energies which are of order $N_c \l^0$.
The $C[f]$ contribution to the interaction reads
\begin{eqnarray}
    V_C =  \sin^2\abs{\,\theta}  V_{C\a}  + \sin^2\abs{\,\theta}\wh{\theta}_1^2  V_{C\b}
    + \sin^2\abs{\,\theta}(\wh{\theta}_2^2 + \wh{\theta}_3^2) V_{C\g} +  \cos^2\abs{\,\theta} V_{C\d} \ , \label{VCgeneral1}
\end{eqnarray}
with
\begin{eqnarray}
   V_{C\a} &\equiv&  N_c b \frac{\wt{\rho}^4}{\wt{d}^2 } \int d^3\wt{x} d\wt{z}\, \wt{z}^2\,\wt{\Box}^2
       \left(\frac{g_- + g_+}{g_- g_+} \right)  \ , \label{Pre1} \\
   V_{C\b} &\equiv& N_c b \frac{\wt{\rho}^4}{\wt{d}^2 } \int \wt{d}^3\wt{x} d\wt{z}\, \wt{z}^2\,\wt{\Box}^2
       \left(\frac{ 4 \wt{z}^2}{g_- g_+} \right)\ , \\
   V_{C\g} &\equiv& N_c b \frac{\wt{\rho}^4}{\wt{d}^2 } \int d^3\wt{x} d\wt{z}\, \wt{z}^2\,\wt{\Box}^2
       \left(\frac{ 4 \wt{x}_2^2}{g_- g_+} \right)\ , \\
   V_{C\d} &\equiv& N_c b \frac{\wt{\rho}^4}{\wt{d}^2 } \int d^3\wt{x} d\wt{z}\, \wt{z}^2\,\wt{\Box}^2
       \left(\frac{ 1}{g_- g_+} \right)\ , \label{Pre4}
\end{eqnarray}
where the cross term in $B^2$ drops by parity and we have rescaled the variable $\wt{x}_M/\wt{d} \ra \wt{x}_M $.
Thus $g_{\pm} \ra \wt{x}_\a^2   + \left(\wt{x}_1 \pm  \frac{1}{2}\right)^2  + {\wt{\rho}^2}/{\wt{d}^2}$. All integrals
are understood in dimensional regularization that preserves both gauge and $O(4)$ symmetry. The
results are
\begin{eqnarray}
V_{C\a} = -V_{C\b} = -V_{C\g} = N_c b \frac{\wt{\rho}^4}{\wt{d}^2 }16\pi^2 \ , \qquad V_{C\d} = 0 \ .
\end{eqnarray}
The $D[f]$ contribution to the interaction reads
%
%
\begin{eqnarray}
V_D \approx  -2 bc N_c  \, \int\,(\,\wt{\Box}^2 \log g_- \,) \,
\inv{\wt{\Box}} \, (\, \wt{\Box}^2 \log g_+ \,) \ .
\label{VD1}
\end{eqnarray}
The Coulomb propagator $1/\wt{\Box}=-1/(4\pi^2 |\wt{x}_+ - \wt{x}_-|^2)$ in 4-dimensions. At large
separations $|\wt{x}_+ - \wt{x}_-|\approx \wt{d}$ and (\ref{VD1}) simplifies to
\begin{eqnarray}
V_D \approx {128\pi^2\,bcN_c}\frac 1{d^2}\,
\left|\frac 1{16\pi^2}\,\int d^3\wt{x} d\wt{z}\, \wt{\Box}^2\log g\right|^2 = \frac{27\pi N_c}{2}\frac 1{\wt{d}^2} \ ,
\label{VD2}
\end{eqnarray}
where the $||$ integrates to the baryon charge 1. $V_D$ captures
the Coulomb repulsion between two unit baryons in 4 dimensions
in the regular gauge. This is not the case in the singular as we
show in Appendix B.

We note that after unscaling $\wt{d} = \sqrt{\l}d$, $V_D\approx N_c/\l$. In regular
gauge, this monopole core repulsion is the Coulomb repulsion between {\it 4-dimensional}
Coulomb charges. We show in Appendix C that this is the natural extension of the
{\it 3-dimensional} omega repulsion at shorter distances in holography.
The repulsion dominates the many-body problem at finite chemical potential as
discussed recently in~\cite{KSZ,SEITZ}. Indeed, for baryonic matter at large baryonic
density $n_B$, the energy is dominated by the Coulomb repulsion (\ref{VD2}). The corresponding
effective interaction is
\begin{eqnarray}
V_{\rm eff}= \frac 12\int\vec{dx}\vec{dy}\left(\phi^+\phi\right)(\vec{x})V_D(\vec{x}-\vec{y})
              \left(\phi^+\phi\right)(\vec{y})\,\ ,
\end{eqnarray}
leading to an energy per volume of order $N_c\,n_B^{5/3}/\l$ as in~\cite{SEITZ}.

\section{Nucleon-Nucleon Interaction: Core}

At large separation, the nucleon-nucleon core interaction
can be readily extracted from the Skyrmion-Skyrmion core interaction
(\ref{VCgeneral1}) as it is linear in the $SO(3)$ rotation $R$. Indeed,
using the standard decomposition~\cite{YAMA}
\begin{eqnarray}
R^{ab} = \frac{1}{3}(R_T^{ab} + \d^{ab} R_S) \ ,
\end{eqnarray}
with
\begin{eqnarray}
R_S = \Tr R \ , \quad  R_T^{ab} = 3 R^{ab} - \d^{ab} \Tr R \ ,
\end{eqnarray}
the spin $R_S$ and tensor $R_T$ contributions respectively, we may decompose
the core potential as
\begin{eqnarray}
V = V_1 + V_S R_S + V_T^{ab}R_T^{ab} \ .
\end{eqnarray}
The scalar $V_1$, spin $V_S$ and tensor $V_T$ contributions
can be unfolded by a pertinent choice of orientations of the
core Skyrmion-Skyrmion interaction. In general,
\begin{eqnarray}
V = V_1 + V_S\left(4\cos^2\!\abs{\,\theta} - 1\right)
+ V_T^{ab}\left[\left(6\hat{\theta}^a \hat{\theta}^b - 2\d^{ab}\right)\sin^2\!\abs{\,\theta}
+ 3\e^{abc}\wh{\theta}^c \sin2\!\abs{\,\theta}\right] \ ,
\end{eqnarray}
after using the SO(3) parametrization (\ref{SO3})
\begin{eqnarray}
R^{ab} = \d^{ab}\cos 2\!\abs{\,\theta} + 2\wh{\theta}^a \wh{\theta}^b \sin^2\!\abs{\,\theta}
+ \e^{abc} \wh{\theta}^c \sin2\!\abs{\,\theta} \ .
\end{eqnarray}

The axial symmetry $V(\theta_1,\theta_2,\theta_3) = V(\theta_1,\theta_3,\theta_2)$
implies that the tensor components of the core satisfy
$V_T^{22} = V_T^{33}, V_T^{12}= V_T^{31},$ and $V_T^{13} = V_T^{21}$. Thus,
$V$ is reduced to

\begin{eqnarray}
V(\theta_1,\theta_2,\theta_3) &=& V_1 + V_S\left(4\cos^2\!\abs{\,\theta} - 1\right)
    + (V_T^{11}-V_T^{22})( 6\hat{\theta}_1^2 - 2) \sin^2\!\abs{\,\theta} \nn \\
  &+& (V_T^{12}+V_T^{13})( 6\hat{\theta}_1(\hat{\theta}_2 + \hat{\theta}_3)  ) \sin^2\!\abs{\,\theta})
    + (V_T^{12}-V_T^{13})( 3 (\hat{\theta}_2 + \hat{\theta}_3) \sin2\!\abs{\,\theta}  ) \nn \\
  &+& (V_T^{23}+V_T^{32})(  6\hat{\theta}_2 \hat{\theta}_3 \sin^2\!\abs{\,\theta} )
    + (V_T^{23}-V_T^{32})( 3\hat{\theta}_1 \sin2\!\abs{\,\theta} ) \ .
\end{eqnarray}
In particular,
\begin{eqnarray}
&& V(0,0,0) = V_1 + 3 V_S \ , \quad  V(0,0,\pi/2) = V_1 - V_S - 2 (V_T^{11} -  V_T^{22}) \ ,  \\
&& V(\pi/2,0,0) = V_1 - V_S + 4 (V_T^{11} -  V_T^{22})  \ ,
\end{eqnarray}
%
so that
\begin{eqnarray}
&& V_1 = \frac{1}{4}\left[V(0,0,0)+ V(0,0,\pi/2) + V(0,\pi/2,0) + V(\pi/2,0,0)     \right] \ ,  \label{V1}\\
&& V_S = \frac{1}{4}\left[V(0,0,0) - \frac{1}{3}\left( V(0,0,\pi/2) + V(0,\pi/2,0) + V(\pi/2,0,0) \right)\right] \ , \\
&& V_T^{11}-V_T^{22} = \frac{1}{6}\left[V(\pi/2,0,0) - V(0,0,\pi/2)   \right] \label{VT} \ .
\end{eqnarray}

\begin{figure}[]
  \begin{center}
  \includegraphics[width=10cm]{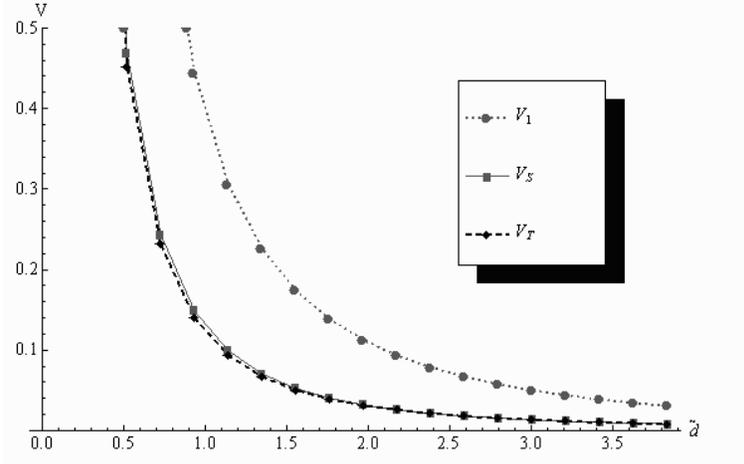}
  \caption{ $V_1,V_S,V_T$ in regular gauge}
  \label{Fig:NNr}
  \end{center}
\end{figure}

Using (\ref{Pre1})-(\ref{Pre4}) we deduce the scalar,
spin and tensor core contributions in the form
\begin{eqnarray}
&& V_1 
= \frac{1}{4}\left(3 V_{C\a} + V_{C\b} + 2 V_{C\g} + V_{C\d} \right) + V_{D} = V_D \ , \nn \\
&& V_{S} = 
\frac{1}{4} \left( - V_{C\a} - \frac{1}{3}V_{C\b} - \frac{2}{3}V_{C\g} + V_{C\d}   \right) = 0 \ , \nn \\
&& V_T^{11}-V_T^{22} = \frac{1}{6} (V_{C\b} - V_{C\g}) = 0 \ .      \label{NNV}
\end{eqnarray}
The off-diagonal tensor $V_T$ core contribution vanishes. This is clear from (\ref{VCgeneral1}).
Indeed (\ref{VCgeneral1}) can be decomposed as
\begin{eqnarray}
    V &=&  \sin^2\abs{\,\theta}  V_{C\a}  + \sin^2\abs{\,\theta}\wh{\theta}_1^2  V_{C\b}
    + \sin^2\abs{\,\theta}(\wh{\theta}_2^2 + \wh{\theta}_3^2) V_{C\g} +  \cos^2\abs{\,\theta} V_{C\d}  + V_D \nn \\
      &=&  \sin^2\abs{\,\theta}  (V_{C\a} + V_{C\g})   + \sin^2\abs{\,\theta}\wh{\theta}_1^2  (V_{C\b} - V_{C\g})
    +  \cos^2\abs{\,\theta} V_{C\d}  + V_D \nn \\
    &=& \ \ \frac{1}{4}\left(3 V_{C\a} + V_{C\b} + 2 V_{C\g} + V_{C\d} \right) + V_{D}  \nn \\
    & & + \frac{1}{4} \left( - V_{C\a} - \frac{1}{3}V_{C\b} - \frac{2}{3}V_{C\g}
+ V_{C\d}   \right) \left(4\cos^2\!\abs{\,\theta} - 1\right) \nn \\
    & & + \frac{1}{6} (V_{C\b} - V_{C\g})( 6\hat{\theta}_1^2 - 2) \sin^2\!\abs{\,\theta} \ ,
\end{eqnarray}
in agreement with (\ref{NNV}). In summary
\begin{eqnarray}
 V_1 = V_D \ ,
\end{eqnarray}
and all others vanish.
For general distances,  we plot $V_1,V_S$ and $V_T$ with (\ref{V1})-(\ref{VT}) in Fig.(\ref{Fig:NNr})
in the regular gauge. The relative separation $\tilde{d}$ is in units of the core size $\tilde{\rho}=9.64$.

\begin{figure}[]
  \begin{center}
  \includegraphics[width=10cm]{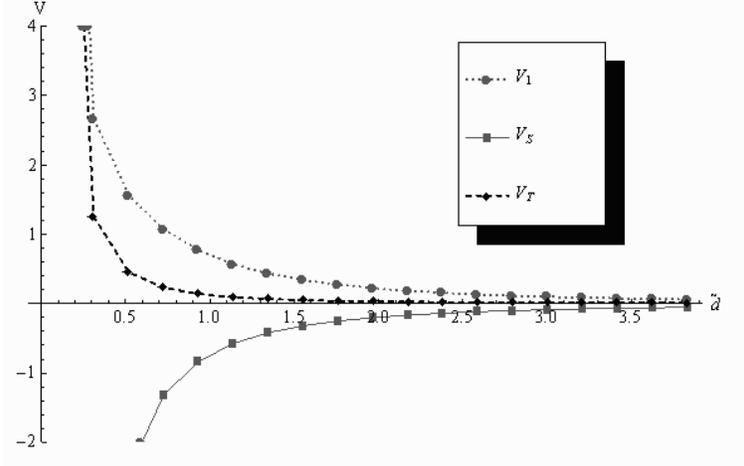}
  \caption{$V_1,V_S,V_T$ in singular gauge }
  \label{Fig:NNs}
  \end{center}
\end{figure}

In the singular gauge, the $V_C$ core contribution to the nucleon-nucleon
interaction remains unchanged while the $V_D$ contribution changes. As a
result, the spin and tensor channels remain the same for both regular and
singular gauges. The central or scalar channel $V_1=V_D$ changes from repulsive
$N_c/\l \wt{d}^2$ (regular) to attractive $-N_c/\l^4 \wt{d}^8$ (singular)
asymptotically. The flip is from monopole to dipole as we detail in Appendix
B. The short distance repulsion in the regular gauge is the 4-dimensional
extension of the 3-dimensional omega repulsion. In~Fig.(\ref{Fig:NNs}) we
show $V_1,V_S$ and $V_T$ with (\ref{V1})-(\ref{VT}) and (\ref{Extra})
in the singular gauge.

\section{Nucleon-Nucleon Interaction: Cloud}

To assess the nucleon-nucleon interaction beyond the core
contribution we need to do a semiclassical expansion around
the $k=2$ configuration, thereby including the effects of
pions and vector mesons as quantum fluctuations around
the core. We refer to these contributions as the cloud.
The semiclassical expansion around the $k=2$ configuration
parallells entirely the same expansion around
the $k=1$ instanton as detailed in~\cite{FF}. The
expansion will be carried out in the axial gauge $A_z=0$
for the fluctuations. This gauge has the merit of exposing
explicitly the pion-nucleon coupling. All cloud calculations
will be carried with the background $k=2$ instanton in the
regular gauge. Some of the results in the singular gauge
are reported in Appendix C.

\subsection{Pion}

In the axial gauge $A_z=0$ for the fluctuations, the pion
coupling to the flavor instanton is explicit in bulk. Indeed,
following the general expansion in~\cite{FF} we have for the
pion-instanton linear coupling
\begin{eqnarray}
S = -\k \int d^4x dz \dell_z \left(K F_{z\m}^a C^{\m,a}\right) \ ,
\label{LINEARVERTEX}
\end{eqnarray}
with the explicit pion field
\begin{eqnarray}
C^{\m,a} \equiv \frac{1}{f_{\pi}}\dell^\m \Pi^a \psi_0 \ , \quad \psi_0 = \frac{2}{\pi}\arctan z \ ,
\end{eqnarray}
and $f_\pi=4\k/\pi$.
As noted in~\cite{FF} all linear meson couplings to the
flavor instanton are boundary-like owing to the soliton
character of the $k=2$ instanton. Since $K F_{zi} \psi_0$
is odd in $z$, for a static instanton,
\begin{eqnarray}
S = \k \int d^4x F_{zi}^a \,K \psi_0 \Big|_{B} \frac{\dell_i\Pi^a}{f_\pi} \ .
\label{PIVERTEX}
\end{eqnarray}
Here $B=\pm z_c$ refers to boundary of the core when using
the non-rigid quantization scheme. To avoid double
counting, for $z<z_c$ the mesons are excluded in the
holographic direction. $z_c$ plays the role of the bag
radius. It will be reduced to $z_c\rightarrow 0$
at the end of all calculations, making the non-rigid
quantization constraint point-like.

The linear pion-2-instanton vertex (\ref{PIVERTEX}) contributes to the
energy through second order perturbation. Specifically,
\begin{eqnarray}
V_\Pi &=& \frac{4\k^2K(z_c)^2\psi_0(z_c)^2 }{2\!\,f_\pi^2}  \int d \vec{x} d \vec{y} F_{iz}^a(\vec{x},z_c)
\langle  \dell_i\Pi(\vec{x})^a \dell_j\Pi(\vec{y})^b \rangle F_{jz}^b(\vec{y},z_c) \label{Vpi}  \\
&=& \frac{\k^2K(z_c)^2\psi_0(z_c)^2 }{2\pi f_\pi^2}  \int d \vec{x} d \vec{y}  F_{iz}^a(\vec{x},z_c)
\dell_i\dell_j  \frac{1}{\abs{\vec{x}-\vec{y}}}   F_{jz}^a(\vec{y},z_c) \ ,
\label{factor2}
\end{eqnarray}
for massless pions. At large separations, the field strength $F_{iz}^a$ splits into
two single instantons of relative distance $d$ and flavor orientation $R$. At large
relative separation $d$, (\ref{factor2}) simplifies to
\begin{eqnarray}
V_\Pi \approx \frac{9}{16\pi f_\pi^2} J_{A}^{ai}(0) D_{ij} J^{Raj}_{A}(0) \ ,
\label{POT0}
\end{eqnarray}
with $D_{ij}=(3\hat{d}_i\hat{d}_j-\delta_{ij})/d^3$. The spatial
component of the axial vector current $J_A$ is unrotated while
$J_A^R$ is rotated. From Appendix D, its zero momentum limit
reads
\begin{eqnarray}
J_{A}^{ai}(0)
\equiv  J_{A}^{ai}(\vec{q} = 0) =  - \frac{4}{3} \k K(z_c)
\psi_0(z_c) \int d \vec{x}\, F_{iz}^a(\vec{x},z_c)  \,\,. \label{JA}
\end{eqnarray}
The projected potential $V_\Pi$ yields
\begin{eqnarray}
\langle s_1 t_1 s_2 t_2 | V_\Pi | s_1 t_1 s_2 t_2 \rangle
&=& \frac{9}{16\pi f_\pi^2} \langle s_1 t_1 |  J_{A}^{ai}(0) | s_1 t_1 \rangle
D_{ij}   \langle s_2 t_2 |  J_{A}^{RAj}(0) | s_2 t_2 \rangle \nn \\
&\equiv&  \frac{1}{16\pi} \left(\frac{g_A}{f_\pi}\right)^2 \inv{d^3}
  \left(3 (\vec{\s}_1 \cdot \wh{d} ) (\vec{\s}_2 \cdot \wh{d}) - \vec{\s}_1 \cdot \vec{\s}_2 \right)
  \left(\vec{\t}_1 \cdot \vec{\t}_2 \right) \ ,
\label{POT1}
\end{eqnarray}
where $g_A=32\k \pi\rho^2/3$ is the axial-vector charge of the
nucleon as detailed in Appendix D. $g_A\approx N_c\l^0$ in
hQCD.

In the $A_z=0$ gauge, the
linear pion-2-instanton vertex (\ref{LINEARVERTEX}) yields a
tensor contribution to the nucleon-nucleon potential
\begin{eqnarray}
V_{T,\Pi } = \frac{1}{16\pi} \left(\frac{g_A}{f_\pi}\right)^2 \inv{d^3}\,\,.
\label{TENSOR}
\end{eqnarray}
This is in agreement with the pseudo-vector one-pion exchange potential
\begin{eqnarray}
V_{T,\Pi }=\frac {(g_{\pi NN}/2M)^2}{4\pi}\,\frac 1{d^3} \ ,
\label{NNVT}
\end{eqnarray}
if we identify
\begin{eqnarray}
\frac{g_{\pi NN}}{M_N}=\frac{g_A}{f_\pi}\,\,.
\label{GT}
\end{eqnarray}
This is just the Goldberger-Treiman relation which is also satistified
by the holographic construction in the $A_z=0$ gauge and for massless
pions.

In reaching (\ref{POT1}) and  the relation (\ref{GT}) there is
a subtlety. Indeed in (\ref{POT0}) the pion propagator $D_{ij}$ is
supposed to be longitudinal and the axial vector source $J_A^{ij}$
transverse, so that the contraction vanishes. The subtlety arises
from the ambiguity in the axial vector source at zero momentum and
for massless pions as discussed in Appendix D. The contraction is
ambiguous through $0/0$. The ambiguity is lifted by the order of
limits detailed in Appendix D, which effectively amounts to a
longitudinal component of the axial vector source at zero momentum.
This result is independently confirmed by using the strong coupling
source theory discussed in Appendix C.

Finally, the pion coupling (\ref{PIVERTEX})
in the axial gauge is pseudoscalar and strong
and of order $\sqrt{N_c/\l}$. The reader may object that this
conclusion maybe at odd with naive $1/N_c$ power counting whereby the pseudovector
coupling is of order $\sqrt{N_c}/{N_c}\approx 1/\sqrt{N_c}$ with the
extra $1/N_c$ suppression brought about by the $\gamma_5$ in the
nucleon axial vector source~\cite{YAMA}. In strongly coupled
models such as hQCD the nucleon source is of order $N_c^0$ not $1/N_c$,
and yet chiral symmetry is fully enforced in the nucleon sector.
hQCD is a chiral and dynamical version of the static Chew model of 
the $\D$ for strong coupling~\cite{CHEW}. 
Also, the reader may object that the one-pion iteration which
is producing a potential of order $N_c/\l$, may cause an even
stronger correction by double iteration of order $(N_c/\l)^{3/2}$ and
so on. This does not happen though, since the direct and
crossed diagram to order $(N_c/\l)^{3/2}$ cancel at strong
coupling. The same cancellation is at the origin of unitarization
in $ \pi \, N \ra \pi \, N $ scattering (Bhabha-Heithler mechanism).

\subsection{Axials}

The linear vertex (\ref{LINEARVERTEX}) also couples vector and axial
vector mesons to the 2-instanton solution at the core in bulk. For
instance, the axial-vector meson contribution
follows from (\ref{LINEARVERTEX}) by inserting
\begin{eqnarray}
C^{\m,a} \equiv a_\m^{a,n}\psi_{2n} \ ,
\end{eqnarray}
so that
\begin{eqnarray}
S = 2\k \int d^4x \left(K {\mathbb{F}}^{b,z \m} {a}_\m^{b,n}\psi_{2n} \right) \Big|_{z=z_c}\ .
\label{LINEARVERTEXV}
\end{eqnarray}
The sum over $n$ is subsumed. We have used the fact that $K F_{zi} \psi_{2n}$ is odd in $z$
(axial exchange) and  that the surface contribution at $z=\infty$ is zero since
${{F}^b}_{z\n} \sim \d(z)$ is localized in bulk to leading order in $1/\l$.

In second order perturbation, (\ref{LINEARVERTEXV}) contributes a static potential
\begin{eqnarray}
V_{A}
&=& 2\k^2K(z_c)^2\psi_{2n}(z_c)\psi_{2m}(z_c)   \int d \vec{x} d \vec{y}  F_{iz}^a(\vec{x},z_c)
\D_{ij}^{mn,ab}  F_{jz}^b(\vec{y},z_c) \label{VA1} \ .
\end{eqnarray}
At large separations, the field strength $F_{iz}^a$ splits into
two single instantons of relative distance $d$ and flavor orientation $R=R_1^TR_2$. At large
relative separation $d$, (\ref{VA1}) simplifies to~\footnote{For simplicity we often omit $|_{z=z_c}$ and $\psi_n \equiv \psi_n(z_c)$. }
\begin{eqnarray}
V_A &\approx& \frac{9}{16\pi} \sum_n J_{A}^{ai}(0) \left(\frac{\psi_{2n}}{\psi_0}\right)^2
 \left(-\d_{ij} + \frac{\dell_i \dell_j}{m_{2n}^2}\right) \frac{e^{-m_{2n}d}}{d} J^{Raj}_{A}(0) \nn \\
  &=&  \frac{9}{16\pi} \sum_n J_{A}^{ai}(0) \left(\frac{\psi_{2n}}{\psi_0}\right)^2
     \left[ \left( 1 + \frac{2   }{m_{2n}d}  + \frac{3   }{m_{2n}^2d^2} \right)
\wh{d}_i \wh{d}_j  \right. \nn \\
&& \qquad \qquad \qquad \qquad \qquad \quad  \left.  - \d_{ij}\left(1+\inv{m_{2n}^2d^2}\right) \right] \frac{e^{-m_{2n}d}}{d} J^{Raj}_{A}(0) \ , \label{NNVA}
\end{eqnarray}
where $J_{A}^{ai}(0)$ is defined in (\ref{JA}) and  the spatial component of the axial vector current $J_A$ is unrotated while
$J_A^R$ is rotated. The projected potential $V_A$ yields
\begin{eqnarray}
& & \langle s_1 t_1 s_2 t_2 | V_A | s_1 t_1 s_2 t_2 \rangle \nn \\
& & \qquad \approx \frac{g_A^2}{16\pi}  \sum_n
 \left(\frac{\psi_{2n}}{\psi_0}\right)^2  e^{- m_{2n} d} \left(-\inv{d}-\inv{m_{2n}^2d^3}\right)
  \left( \vec{\s}_1 \cdot \vec{\s}_2 \right) \left(\vec{\t}_1 \cdot \vec{\t}_2 \right) \nn \\
& & \qquad \quad + \frac{g_A^2}{16\pi}  \sum_n
 \left(\frac{\psi_{2n}}{\psi_0}\right)^2  e^{- m_{2n} d} \left( \inv{d} + \frac{2   }{m_{2n}d^2}  + \frac{3   }{m_{2n}^2d^3} \right)
   (\vec{\s}_1 \cdot \wh{d} ) (\vec{\s}_2 \cdot \wh{d})  \left(\vec{\t}_1 \cdot \vec{\t}_2 \right) \nn \\
& & \qquad \approx \frac{g_A^2}{16\pi}  \sum_n
 \left(\frac{\psi_{2n}}{\psi_0}\right)^2  \frac{e^{- m_{2n} d}}{d}
   \left[(\vec{\s}_1 \cdot \wh{d} ) (\vec{\s}_2 \cdot \wh{d})  - \left( \vec{\s}_1 \cdot \vec{\s}_2 \right)    \right]
   \left(\vec{\t}_1 \cdot \vec{\t}_2 \right) \ ,
\end{eqnarray}
which contributes to the spin $V_{S,A}$ and tensor part $V_{T,A}$ of the
NN interaction. Specifically,
\begin{eqnarray}
&& V_{S,A} \approx \sum_n  G_{SA,2n}^2 \frac{e^{- m_{2n}} d}{4\pi\,d} \ ,
  \quad V_{T,A} \approx\sum_n  G_{TA,2n}^2 \frac{e^{- m_{2n}} d}{4\pi\,d} \ , \nn \\
&& G_{SA,2n} \equiv - \frac{g_A\psi_{2n}}{\sqrt{6} \psi_0} \sim  g_A m_{2n}/\sqrt{\kappa}\ , \qquad
   G_{TA,2n} \equiv  \frac{g_A\psi_{2n}}{\sqrt{12} \psi_0} \sim  g_A m_{2n}/\sqrt{\kappa} \ , \label{Axialvector}
\end{eqnarray}
with $G_{SA,2n}\approx \sqrt{N_c/\l}$ and $G_{TA,2n}\approx\sqrt{N_c/\l}$ the spin and tensor couplings
of the tower of axials to the nucleon.

\subsection{Vectors}

For the  vector mesons the time component $F_{0z}$ contribution is leading in $N_c$
compared to the space component $F_{iz}$. This is the opposite of the axial vector
contribution. For the SU(2) part (rho, rho', ...), we have
\begin{eqnarray}
V_{V} &=& \frac 1{2\pi}
\sum_{n}\,\k^2 K(z_c)^2 \psi_{2n-1}^2(z_c)  \int d \vec{x} d \vec{y}  F_{0 z}^a(\vec{x},z_c)
\frac{e^{-m_{2n-1}\abs{\vec{x}-\vec{y}}}}{\abs{\vec{x}-\vec{y}}}    F_{0 z}^a(\vec{y},z_c) \nn \\
&\approx& \frac{1}{4\pi} \sum_n J^{a} \psi_{2n-1}^2
 \frac{e^{-m_{2n-1}d}}{d}  J^{Ra}  \ , \label{NNVV}
\end{eqnarray}
where $J^a \equiv \int d\vec{x} 2\k K F^a_{z0}\Big|_{z=z_c} $ is the unrotated angular momentum
in~\cite{FF}. We note that $R=R^T_1R_2$ and that $R_2^{ab}J^b=-I_2^a$ where $I_2^a$ is the unrotated
isovector charge of the nucleon labelled 2. The same holds for label 1. Thus
\begin{eqnarray}
\langle s_1 t_1 s_2 t_2 | V_V | s_1 t_1 s_2 t_2 \rangle \approx  \sum_n \frac 14 \psi_{2n-1}^2
 \frac{e^{-m_{2n-1}d}}{4\pi\,d}     \left(\vec{\t}_1 \cdot \vec{\t}_2 \right) \ ,
\end{eqnarray}
which is seen to contribute to the isospin part of the central potential
\begin{eqnarray}
V_{1,V}^{-} \approx\sum_n G_{1V,2n-1}^2
 \frac{e^{-m_{2n-1}d}}{4\pi\,d} \label{1V-} \ ,
\end{eqnarray}
with $G_{1V,2n-1}=\psi_{2n-1}/2\approx 1/\sqrt{N_c\l}$. This contribution is subleading
in the potential.

Similarly, the U(1) vector contribution part (omega, omega', ...) reads
\begin{eqnarray}
V_{\wh{V}} &\approx& \frac{N_c^2}{16\pi} \sum_n B \psi_{2n-1}^2
 \frac{e^{-m_{2n-1}d}}{d} B = \frac{N_c^2}{4} \sum_n  \psi_{2n-1}^2
 \frac{e^{-m_{2n-1}d}}{4\pi d}   \ ,  \label{NNVV1}
\end{eqnarray}
where $B \equiv \int d\vec{x} \frac{4}{N_c}\k K \wh{F}_{z0} \Big|_{z=z_c}$ is the baryon number
introduced in \cite{FF}. This contribution to the central potential is leading
\begin{eqnarray}
&&V_{1,\wh{V}} \equiv V_{\wh{V}}   \approx   \sum_n G^2_{\wh{V},2n-1}
 \frac{e^{-m_{2n-1}d}}{4\pi\, d}   \ , \label{1V+} \\
&& G_{\wh{V},2n-1} \equiv \frac{N_c}{2} \psi_{2n-1} \ ,
\end{eqnarray}
with $G_{\wh{V},2n-1}\approx \sqrt{N_c/\l}$.

For completeness, we quote the spatial contributions from the vectors,
both of which are subleading in the potential. The $SU(2)$ vector meson
contribution is
\begin{eqnarray}
V_{V}'
&=& 2\k^2K(z_c)^2\psi_{2n-1}(z_c)\psi_{2m-1}(z_c)   \int d \vec{x} d \vec{y}  F_{iz}^a(\vec{x},z_c)
\D_{ij}^{mn,ab}  F_{jz}^b(\vec{y},z_c) \ .\label{VV1}
\end{eqnarray}
At large separations, the field strength $F_{iz}^a$ splits into
two single instantons of relative distance $d$ and flavor orientation $R=R_1^TR_2$. At large
relative separation $d$, (\ref{VA1}) simplifies to
\begin{eqnarray}
V_{V}' &\approx& \frac{9}{16\pi} \sum_n J_{V}^{ai}(0) \left(\psi_{2n-1}\right)^2
 \left(-\d_{ij} + \frac{\dell_i \dell_j}{m_{2n-1}^2}\right) \frac{e^{-m_{2n-1}d}}{d} J^{Raj}_{A}(0) \nn \\
  &=&  \frac{9}{16\pi} \sum_n J_{V}^{ai}(0) \left(\psi_{2n-1}\right)^2
     \left[ \left( 1 + \frac{2   }{m_{2n-1}d}  + \frac{3   }{m_{2n-1}^2d^2} \right)
\wh{d}_i \wh{d}_j \right. \nn \\
&& \qquad \qquad \qquad \qquad \qquad \quad
\left. - \d_{ij}\left(1+\inv{m_{2n-1}^2d^2}\right) \right] \frac{e^{-m_{2n-1}d}}{d} J^{Raj}_{V}(0) \ , \label{NNVA}
\end{eqnarray}
where $J_{V}^{ai}(0) \equiv -(4/3) \k K \int d\vec{x} F_{iz}^a $ and  the spatial component of the vector current $J_V$ is unrotated while
$J_V^R$ is rotated. The projected potential $V_V'$ yields
\begin{eqnarray}
& & \langle s_1 t_1 s_2 t_2 | V_V' | s_1 t_1 s_2 t_2 \rangle \nn \\
& & \qquad \approx \frac{g_V^2}{16\pi}  \sum_n
 \left(\psi_{2n-1}\right)^2  e^{- m_{2n-1} d} \left(-\inv{d}-\inv{m_{2n-1}^2d^3}\right)
  \left( \vec{\s}_1 \cdot \vec{\s}_2 \right) \left(\vec{\t}_1 \cdot \vec{\t}_2 \right) \nn \\
& & \qquad \quad + \frac{g_V^2}{16\pi}  \sum_n
 \left(\psi_{2n-1}\right)^2  e^{- m_{2n-1} d} \left( \inv{d} + \frac{2   }{m_{2n-1}d^2}  + \frac{3   }{m_{2n-1}^2d^3} \right)
   (\vec{\s}_1 \cdot \wh{d} ) (\vec{\s}_2 \cdot \wh{d})  \left(\vec{\t}_1 \cdot \vec{\t}_2 \right) \nn \\
& & \qquad \approx \frac{g_V^2}{16\pi}  \sum_n
 \left(\psi_{2n-1}\right)^2  \frac{e^{- m_{2n-1} d}}{d}
   \left[(\vec{\s}_1 \cdot \wh{d} ) (\vec{\s}_2 \cdot \wh{d})  - \left( \vec{\s}_1 \cdot \vec{\s}_2 \right)    \right]
   \left(\vec{\t}_1 \cdot \vec{\t}_2 \right) \ ,
\label{POTVPRIME}
\end{eqnarray}
with $J_V^{ai}(0)=g_V\delta^{ai}$. (\ref{POTVPRIME}) contributes to both the spin
$V_{S,V}'$ and tensor part $V_{T,V}'$ of the NN interaction,
\begin{eqnarray}
&& V'_{S,V} \approx \frac{1}{4\pi} \sum_n  G_{SV,2n-1}^2 \frac{e^{- m_{2n-1} d} }{d} \ ,
  \quad V'_{T,V} \approx \frac{1}{4\pi} \sum_n  G_{TV,2n-1}^2 \frac{e^{- m_{2n-1} d} }{d} \ , \nn \\
&& G_{SV,2n} \equiv - \frac{g_V\psi_{2n-1}}{ \sqrt{6} } \ , \qquad \qquad \qquad
   G_{TV,2n} \equiv  \frac{g_V\psi_{2n-1}}{\sqrt{12} } \ . \label{Vector1}
\end{eqnarray}
The holographic description of the nucleon-nucleon
potential is consistent with the meson-exchange potentials in
nuclear physics. Holography allows a systematic organization of the
NN potential in the context of semiclassics, with the NN interaction
of order $N_c/\l$ in leading order.

\section{Holographic NN potentials}

In general, the NN potential in holography is composed of the core and the cloud contributions to order $N_c/\l$.
For non-asymptotic distances, both the core and cloud contributions have a non-linear
dependence on the rotation matrix $R(U)$, making the projection on the NN channel involved.
Formally, the potential (core plus cloud) can be expanded using the irreducible representations
of SU(2). Specifically
\begin{eqnarray}
V(d,U)=\sum_{j=0}^\infty\sum_{m=-j}^{+j}\,V_{jm}(d) \,D_{mm}^j(U) \ ,
\label{IRREP}
\end{eqnarray}
where $D_{m,m'}^j(U)$ are U-valued Wigner functions. For $k=2$ the azimuthal symmetry restrics $m'=m$
with $V_{jm}=V_{j,-m}$. In particular
\begin{eqnarray}
V_{jm}(d)=\frac{2\pi^2}{(2j+1)}\int\,dU\,V(d,U)\,D^{j\ *}_{m,m}(U)\,\,.
\end{eqnarray}
The projection on the NN channel follows by sandwiching (\ref{IRREP}) between the
normalized NN states $D^{1/2}(1)\otimes D^{1/2}(2)$. While straightforward, this
procedure is involved owing to the complicated nature of the $k=2$ instanton both
in the core and in the cloud on $R(U)$. In general
\begin{eqnarray}
V(d,U) = V_{\mathrm{core}}(d,U) + V_{\mathrm{cloud}}(d,U) \ .
\end{eqnarray}
$V_{\mathrm{core}}(d,U)$ is defined as
\begin{eqnarray}
&& V_{\mathrm{core}}(d,U) \equiv  \D E[f_{-+}] - \D E[f_-] - \D E[f_+] \ , \nn \\
&& \qquad  \D E[f]  = \frac{\k}{6\l} \int d^3\wt{x} d\wt{z} \left(\wt{z}^2-\frac{3^7
\pi^2}{2^4} \wt{\Box} \log \abs{f} \right) \wt{\Box}^2 \log \abs{f} \ ,
\end{eqnarray}
where
\begin{eqnarray}
 && \log \abs{f_{-+}} \equiv  -  \log \left[ \left( g_- +
 A \right) \left( g_+  + A  \right)  - B^2 \right] \ ,  \\
 && \log \abs{f_{\pm}} =  -  \log g_{\pm} \ ,  \\
 &&   g_{\pm} =  \sum_{\a=2,3,4}\wt{x}_\a^2   + \left(\wt{x}_1 \pm  \frac{\wt{d}}{2}\right)^2  + \wt{\rho}^2 \ , \qquad
 A = \frac{\wt{\rho}^4 \sin^2\abs{\,\theta}}{\wt{d}^2} \ ,  \\
 &&   B =  \wt{\rho}^2 \left(\cos\abs{\,\theta} + \frac{2}{\wt{d}} \sin\abs{\,\theta}
  \left[\wh{\theta}_1 \wt{z} +\wh{\theta}_2 \wt{x}_3 - \wh{\theta}_3 \wt{x}_2  \right] \right) \ .
\end{eqnarray}
$V_{\mathrm{cloud}}(d,U)$ is defined as
\begin{eqnarray}
&& V_{\mathrm{cloud}}(d,U) \equiv 2\k^2 K^2 \sum_{n}\psi_n^2 \int d\vec{x} d\vec{y}  \\
&& \quad \Big[ F_{iz}^a(x_M;-+)\D^{ij}_n(\vec{x}-\vec{y})\,F_{iz}^a(y_M;-+)
+ \wh{F}_{0z}(x_M;-+)\D^{00}_n(\vec{x}-\vec{y}) \wh{F}_{0z}(y_M;-+) \nn \\
&& \quad \ \
- 2{F}_{iz}^a(x_M;-)\D^{ij}_n(\vec{x}-\vec{y})\,{F}_{jz}^a(y_M;-)
 - 2\wh{F}_{0z}(x_M;-)\D^{00}_n(\vec{x}-\vec{y}) \wh{F}_{0z}(y_M;-) \Big]\Big|_{z=z_c} \ , \nn
\end{eqnarray}
where $F_{iz}^a(x_M;-)$ and $F_{iz}^a(x_M;-+)$ are the field strengths of the
$k=1$ and the $k=2$ $SU(2)$ instanton respectively,
\begin{eqnarray}
&& F_{iz}^a(x_M;-) = -2 \d_{ai} \t^{a} \frac{\rho^2}{(\xi_-^2 + \rho^2)^2} \ ,  \\
&& F_{iz}^a(x_M;-+) = - 2 \d_{ai}U^\dg \mathbb{B}\left( f_{-+} \otimes \t^a \right) \mathbb{B}^\dg U \ ,  \\
&& \qquad \mathbb{B}^\dg = \begin{pmatrix}
      0 & -\mone & 0\\
      0 & 0 & -\mone \\
    \end{pmatrix} \ , \qquad f^{-1}_{-+} = \begin{pmatrix}
  g_- + A & B \\
  B & g_+  + A \\
   \end{pmatrix}   \ ,   \nn  \\
&& \qquad g_{\pm} \equiv x_\a^2   + \left(x_1 \pm  \frac{d}{2}\right)^2
+ \rho^2 \ , \quad x_\a^2 \equiv  x_2^2 + x_3^2 + x_4^2  \ , \nn \\
&& \qquad  A \equiv \frac{\rho^4 \sin^2\abs{\,\theta}}{d^2} \ , \quad B \equiv
\rho^2 \left(\cos\abs{\,\theta} + \frac{2}{d} \sin\abs{\,\theta}
  \left[\wh{\theta}_1 z +\wh{\theta}_2 x_3 - \wh{\theta}_3 x_2  \right] \right) \ ,\nn \\
&&\qquad \begin{pmatrix}
                 \l_1^\dg & \frac{d}{2}\t^1 -x^\dg  &  \frac{\rho^2}{d} \sin\!\abs{\,\theta} \wh{\theta}_a \t^a \t^1 \  \\
                  \l_2^\dg &  \frac{\rho^2}{d} \sin\!\abs{\,\theta} \wh{\theta}_a \t^a \t^1 \  & -\frac{d}{2}\t^1 -x^\dg \\
               \end{pmatrix}  U = 0 \ , \qquad  U^\dg U = \mone \ .  \nn
\end{eqnarray}
The U(1) fields $\wh{F}_{z0}(x_M;-+) = \dell_z \wh{A}_0(x_M;-+)$ and
$\wh{F}_{z0}(x_M;-) = \dell_z \wh{A}_0(x_M;-)$ follow from
\begin{eqnarray}
\wh{A}_0(x_M;-+) = \frac{1}{32\pi^2 a} \Box \log \abs{f_{-+}} \ , \qquad \wh{A}_0(x_M;-) = \frac{1}{32\pi^2 a} \Box \log \abs{f_{-}} \ .
\end{eqnarray}
The propagators are defined as
\begin{eqnarray}
\D^{ij}_n(\vec{x}-\vec{y}) = (-\d_{ij}+ \wh{\dell}_i \wh{\dell}_j)
\frac{e^{-m_n \abs{\vec{x}-\vec{y}}}}{4\pi \abs{\vec{x}-\vec{y}}  } \ , \qquad
 \D^{00}_n(\vec{x}-\vec{y})= \frac{e^{-m_n \abs{\vec{x}-\vec{y}}}}{4\pi \abs{\vec{x}-\vec{y}}  } \ .
\end{eqnarray}

If we were to saturate (\ref{IRREP}) by $j=0,1$ which is exact asymptotically as
we have shown both for the core and cloud, then the projection procedure is much
simpler. In particular, the NN potential simplifies to
\begin{eqnarray}
&& V_{NN} = V^+_1 + \vec{\t}_1 \cdot \vec{\t}_2 \, V_1^-
+\vec{\s_1}\cdot\vec{\s_2} \left(V_S^+ + \vec{\t}_1\cdot\vec{\t}_2 \, V_S^- \right) \\
&& \qquad \quad +
\left(3 (\vec{\s}_1 \cdot \wh{d} ) (\vec{\s}_2 \cdot \wh{d}) - \vec{\s}_1 \cdot \vec{\s}_2 \right)
 \left(V_T^+ + \vec{\t}_1\cdot\vec{\t}_2 \, V_T^- \right) \ ,
\end{eqnarray}
with the core contributions
\begin{eqnarray}
&& V^+_{1,\mathrm{core}} = \frac{1}{4}\left[V(0,0,0)+ 2 V(0,0,\pi/2)+ V(\pi/2,0,0)     \right] \ , \\
&& V^-_{S,\mathrm{core}} = \frac{1}{4}\left[V(0,0,0) - \frac{2}{3} V(0,0,\pi/2) -\frac{1}{3} V(\pi/2,0,0) \right] \ , \\
&& V^-_{T,\mathrm{core}} = V_T^{11}-V_T^{22} = \frac{1}{6}\left[V(\pi/2,0,0) - V(0,0,\pi/2)   \right] \ ,
\end{eqnarray}
as detailed above. The cloud contributions $V_1$, $V_S$ and $V_T$ remain the same.
At large distances $d$ the core contribution is dominant and repulsive in the regular
gauge (\ref{VD2})
\begin{eqnarray}
V_{1,\mathrm{core}}^+ \approx \frac{27\pi N_c}{2\l}\frac 1{d^2} \ ,
\end{eqnarray}
and subdominant and attractive in the singular gauge (\ref{VD4})
\begin{eqnarray}
V_{1,\mathrm{core}}^+ \approx  - \frac{81 \pi N_c}{\l^4}\frac {\rho^6}{d^8} \ .
\end{eqnarray}
The dominant cloud contributions are
\begin{eqnarray}
&&V_{1,\wh{V}}^+   \approx   \sum_n G^2_{1\wh{V},2n-1}
 \frac{e^{-m_{2n-1}d}}{4\pi\, d}   \ ,  \qquad  G_{1\wh{V},2n-1} \equiv \frac{N_c}{2} \psi_{2n-1} \ \sim \ \sqrt{\frac{N_c}{\l}} \ ,   \\
&& V_{S,A}^- \approx  \sum_n  G_{SA,2n}^2 \frac{e^{- m_{2n}} d}{4\pi d} \ , \qquad \quad \ \
G_{SA,2n} \equiv - \frac{g_A\psi_{2n}}{\sqrt{6} \psi_0}\quad \  \sim \  \sqrt{\frac{N_c}{\l}}\ ,  \\
&& V_{T,A}^- \approx  \sum_n  G_{TA,2n}^2 \frac{e^{- m_{2n}} d}{4\pi d}  \ , \qquad \quad \ \
G_{TA,2n} \equiv  \frac{g_A\psi_{2n}}{\sqrt{12} \psi_0}\quad \ \  \sim \  \sqrt{\frac{N_c}{\l}}  \ , \\
&& V_{T,\Pi}^- \approx \frac{1}{16\pi} \left(\frac{g_A}{f_\pi}\right)^2 \inv{d^3}\ \ \sim \  \frac{N_c}{\l} \ .
\end{eqnarray}
from (\ref{TENSOR}), (\ref{Axialvector}), (\ref{1V+}), and (\ref{1V-}).
To order $N_c/\l$ we note that $V_1^-=V_S^+=V_T^+=0$.

\section{Conclusions}

We have extended the holographic description of the nucleon suggested in
\cite{Sakai3} to the two nucleon problem. In particular, we have shown how
the exact $k=2$ ADHM instanton configuration applies to the NN problem.
The NN potential is divided into a short distance core contribution and
a large distance cloud contribution that is meson mediated. This is a
first principle description of meson exchange potentials sucessfully
used for the nucleon-nucleon problem in pre-QCD~\cite{BROWN}.

The core contribution in the regular gauge is of order $N_c/\l$. It is
Coulomb like in the central channel. Remarkably, the
repulsion is 4-dimensional Coulomb, a hallmark of holography. This
repulsion dominates the high baryon density problem in holography
as discussed recently in~\cite{KSZ,SEITZ}. The dominant Coulomb repulsion
is changed to subdominant dipole attraction for instantons in the singular
gauge.

We have shown in the context of semiclassics, that the meson-instanton
interactions in bulk is strong and of order $\sqrt{N_c/\l}$. In the
Born-Oppenheimer approximation they contribute to the potentials to
order $N_c/\l$. These cloud contributions dominate at large distances.
The central potential is dominated by a tower of omega exchanges, the
tensor potential by a tower of pion exchanges, while the spin and
tensor potentials are dominated by a tower of axial-vector exchanges.
The vector exchanges are subdominant at large $N_c$ and strong coupling.
Holography, fixes the potentials at intermediate and short distances
without recourse to {\it adhoc} form factors~\cite{BROWN} or truncation as
in the Skyrme model~\cite{ZAHEDBROWN}.

The present work provides a quantitative starting point for an analysis
of the NN interaction in strong coupling and large $N_c$. For a realistic
comparison with boson exchange models, we need to introduce a pion mass.
It also offers a systematic framework for discussing the deuteron
problem, NN form factors and NN-meson  and NN-photon emissions in the
context of holography. We plan to address some of these issues next.

\section{Acknowledgments}

We thank Pierre Van Baal, Tamas Kovacs, Larry McLerran and Sang-Jin Sin
for discussions. This work was supported in part by US-DOE grants
DE-FG02-88ER40388 and DE-FG03-97ER4014.

\appendix

\section{Instantons in Singular gauge}

The $k=1$ instanton
in the singular gauge follows from (\ref{AM}) through a gauge transformation
$g^{-1}=\wh{\xi}=\xi/|\xi|$ which is singular at $\xi=x-X=0$. This is achieved through the
shift $U\rightarrow Ug$, which amounts in general to the new inverse potential
$1/f=1+\rho^2/\xi_M^2$. The corresponding action density is
\begin{eqnarray}
\Tr F_{MN}^2 = \Box^2 \log f \  = \frac{96\rho^4}{((x_M-X_M)^2+\rho^2)^4} -16\pi^2\delta (x_M-X_M)\,\,.
\label{Sonedensity}
\end{eqnarray}
The instanton in the singular gauge is threaded by an antiinstanton of zero
size at its center. Its topological charge is strictly speaking zero. It is
almost 1 if the center $x=X$ is excluded. This point is usually  subsumed.
Singular instantons have more localized gauge fields than regular instantons.

The ADHM solution for $k=2$ in singular gauge is not known. Following
the $k=1$ argument, we may seek it from the regular gauge by applying
a doubly singular gauge transformation
\begin{eqnarray}
g^{-1}\equiv g_+^{-1}\, g_-^{-1}= \wh{\xi}_+\,\wh{\xi}_- \ ,
\label{DOUBLE}
\end{eqnarray}
which is singular at the centers $\xi_\pm=x\pm D=0$ in quaternion
notations. This amounts to shifting $U\rightarrow Ug$ in the ADHM
construction. We guess that (\ref{DOUBLE}) yields the new inverse
potential
\begin{eqnarray}
f^{-1}\rightarrow \frac{f^{-1}}{(|x-D||x+D|)}\,\,, \label{singular}
\end{eqnarray}
in the singular gauge. As a result, the instanton topological charge is
\begin{eqnarray}
\Tr F_{MN}^2 = \Box^2 \log f \rightarrow \Box^2 \log f
               -16\pi^2\left( \delta (\xi_{+M})+\delta (\xi_{-M})\right)\,\,.
\label{Sonedensity}
\end{eqnarray}
The $k=2$ ADHM density is now threaded by two singular anti-instantons
at $\xi_\pm=0$. Singular instantons have more localized gauge fields
$A_M$ than regular instantons. While this point is of relevance for
gauge variant quantitities, it is irrelevant for gauge invariant
quantities with the exception of the topological charge. This point
is important for the central nucleon-nucleon potential as we now
explain.

\section{Core in Singular gauge}

In the singular gauge we substitute $\abs{f}$ as (\ref{singular}), which results in
\begin{eqnarray}
&& \Box \log \abs{f} \ra \Box \log \abs{f} + \frac{4}{\wt{\xi}_+^2} + \frac{4}{\wt{\xi}_-^2} \ , \\
&& \Box^2 \log \abs{f} \ra \Box^2 \log \abs{f}  -16\pi^2 \left( \delta (\wt{\xi}_{+M})+\delta (\wt{\xi}_{-M})\right) \ .
\end{eqnarray}
This gives extra contributions in addition to the result in regular gauge after the subtraction of the self-energy
\begin{eqnarray}
\D E &\ra& \D E + \D E_s \ , \nn \\
\D E_s &\equiv& \left(\frac{\k}{6\l}\right)\left(\frac{3^7  \pi^2}{2^4} \right) \int d^3 \wt{x} d\wt{z} \nn \\
& & \Big[  32\pi^2 \big\{ \wt{\Box} \log \abs{f_\pm} \left( \delta \wt{\xi}_{+M})+\delta (\wt{\xi}_{-M})\right)
          - \wt{\Box} \log \abs{f_+} \delta (\wt{\xi}_{+M}) -  \wt{\Box} \log \abs{f_-} \delta (\wt{\xi}_{-M}) \big\} \nn  \\
& &  \left.  + 16\pi^2 \left( \delta (\wt{\xi}_{-M})\frac{4}{\wt{\xi}_+^2}
+  \delta (\wt{\xi}_{+M})\frac{4}{\wt{\xi}_-^2} \right)  \right] \ ,
\end{eqnarray}
which comes from the second term in (\ref{DE}) while the first term in (\ref{DE}) remains the same.
$f\pm , f_+,$ and $f_-$ are short for the expressions in the regular gauge in (\ref{fpm1})-(\ref{fpm2}).
Thus
\begin{eqnarray}
\D E_s &=& \frac{N_c 27 \pi }{8} \left[ \int d^3 \wt{x} d\wt{z} \Big[ \wt{\Box}
\log \abs{f} \left( \delta (\wt{\xi}_{+M})+\delta (\wt{\xi}_{-M})\right)
\Big] + \frac{16}{\wt{\rho}^2}+ \frac{4}{\wt{d}^2}\right] \nn \\
&=& \frac{N_c 27 \pi }{2} \left\{\frac{4}{\wt{\rho}^2}+ \frac{1}{\wt{d}^2} \   \right.\nn \\
&&  -\ 2 \wt{d}^2 \Big[ \wt{d}^6(2\wt{d}^4 + 5\wt{d}^2\wt{\rho}^2 + 4\wt{\rho}^4) + \wt{d}^4\wt{\rho}^2 \cos^2\abs{\theta}(-2\wt{d}^4
- 4\wt{d}^2\wt{\rho}^2 - 3\wt{\rho}^4\cos(2\abs{\theta}) ) \nn
\end{eqnarray}
\begin{eqnarray}
&&  \quad +2\wt{d}^4\wt{\rho}^2(\wt{d}^4+4\wt{d}^2\wt{\rho}^2 + 5\wt{\rho}^4)\sin^2\abs{\theta}
+ \wt{d}^2\wt{\rho}^6(3\wt{d}^2 + 8 \wt{\rho}^2)\sin^4\abs{\theta} + 2\wt{\rho}^{10}\sin^6\abs{\theta}    \Big] \nn \\
&&  \left.  \div\ \wt{\rho}^2 \Big[ \wt{d}^4(\wt{d}^2+\wt{\rho}^2) - \wt{d}^4\wt{\rho}^2\cos^2\abs{\theta}
+ \wt{d}^2 \wt{\rho}^2 (\wt{d}^2 + 2\wt{\rho}^2) \sin^2\abs{\theta} + \wt{\rho}^6 \sin^4\abs{\theta}   \Big]^2  \right\} \label{Extra} \\
&=& -\left( \frac{N_c 27 \pi }{2\l} \right) \frac{1 + 4\cos(2\abs{\theta})}{{d}^2} +  \calo({d}^{-4}) \qquad ({d} \gg 1) \ .
\end{eqnarray}
For large $d$, the monopole contribution in (\ref{Extra}) is cancelled by the monopole
contribution (\ref{VD2}) in the regular gauge. This cancellation leads to a dipole attraction
in the singular gauge.

The net dipole attraction in the singular gauge is best seen by noting
that (\ref{VD1}) now reads
\begin{eqnarray}
V_D \approx  -2 bc N_c \int  \, \left(\,\wt{\Box}^2 \log \left(1+\wt{\rho}^2/\wt{\xi}_+^2\right) \,\right) \,
\inv{\wt{\Box}} \, \left(\, \wt{\Box}^2 \log  \left(1+\wt{\rho}^2/\wt{\xi}_-^2\right) \right) \ ,
\label{VD3}
\end{eqnarray}
where $\wt{x}_\pm$ refers to the shifted instanton positions. For large separations
$\wt{d}/\wt{\rho}\gg 1$, the leading contribution to $V_D$ is
\begin{eqnarray}
V_D\approx -768\pi^2\,bc\,N_c\,\frac {\wt{\rho}^6}{\wt{d}^8} = - 81 \pi N_c \frac {\wt{\rho}^6}{\wt{d}^8} \,\, ,
\label{VD4}
\end{eqnarray}
by repeated use of the 4-dimensional formulae
\begin{eqnarray}
{\Box}\,\frac 1{\xi^{2n}}=-4\pi^2\,\delta_{n1}\,\delta^4(\xi) + 2n(2(n+1)-4)\frac 1{\xi^{2(n+1)}} \ .
\end{eqnarray}
This contribution is of order $N_c/\l^4$ following the unscaling of
$\wt{d} = \sqrt{\l}d$. (\ref{VD4}) is dipole-like
and attractive. As expected, the threading antiinstanton in the singular
gauge cancels the leading $N_c/\l$ repulsive monopole contribution in
4-dimensional Coulomb's law, resulting in the attractive dipole-like
contribution (van der Waals).

\section{Strong Source Theory}

In this Appendix, we check our cloud calculation in the singular gauge
using the strong coupling source theory used for small cores in~\cite{CHEW,JOHN}
and more recently in holography in~\cite{Hashimoto}. This method provides
an independent check on our semiclassics in the $k=2$ sector.

The energy in the leading order of $\l$ is
\begin{eqnarray}
 E  = \k \ \Tr \int d^3x dz  \left(\half K^{-1/3} F_{ij}^2 +  K  F_{iz}^2 \right)
    + \frac{\k}{2} \ \int d^3x dz  \left(\half K^{-1/3} \wh{F}_{ij}^2 +  K  \wh{F}_{iz}^2 \right) \ .
\end{eqnarray}
where, in the region $1 \ll \xi$,
\begin{eqnarray}
 && \wh{A}_0 \approx -\frac{1}{2a\l}(G_- + G_+) \ ,  \\
 && A_i^a \approx -2\pi^2\rho^2 \left( \left( \e^{iaj}\dell_{+j} - \d^{ia}\dell_{+Z} \right) G_+
 + R^{ab}\left( \e^{ibj}\dell_{-j} - \d^{ib}\dell_{-Z}\right) G_-  \right) \ ,  \\
 && A_z^a \approx -2\pi^2\rho^2 \left(\dell_{+a} H_+  + R^{ab} \dell_{-b} H_-  \right) \ ,
\end{eqnarray}
with
\begin{eqnarray}
  && G_{\pm} \approx \k \sum_{n=1}^{\infty}  \psi_n(z) \psi_n(\pm Z) Y_n \left( \big|\vec{x} - \vec{X}_\pm \big| \right)  \ ,  \\
  && H_{\pm} \approx \k \sum_{n=0}^{\infty}  \phi_n(z) \phi_n(\pm Z) Y_n \left( \big|\vec{x} - \vec{X}_\pm \big| \right)  \ ,  \\
  && \phi_0(z) \equiv \frac{1}{\sqrt{\k \pi} K(z)} \ , \quad \phi_n(z) =
\frac{1}{\sqrt{\l_n}} \dell_z \psi_n(z) \quad (n=1,2,3,\cdots) \ ,  \\
  && Y_n \left( \big|\vec{x} - \vec{X}_\pm \big| \right)  \equiv -\frac{e^{-\sqrt{\l_n}
\big|\vec{x} - \vec{X}_\pm \big| }}{4\pi \big|\vec{x} - \vec{X}_\pm \big| } \ ,
\quad \dell_{\pm a} \equiv \frac{\dell}{\dell X_\pm^a}  \ , \quad \dell_{\pm Z} \equiv \frac{\dell}{\dell Z_\pm} \ .
\end{eqnarray}

\subsection{Pion}

The pion contribution stems from
\begin{eqnarray}
 E_{\Pi}  = \frac{\k}{2} \int d^3x dz K \left(\dell_i A_z^a\right)\left( \dell_i A_z^a\right) \ ,
\end{eqnarray}
where
\begin{eqnarray}
 && A_z^a \approx -2\pi^2\rho^2 \left(\dell_{+a} H_+  + R^{ab} \dell_{-b} H_-  \right) \ , \label{Azlong}
\end{eqnarray}
with $\phi_0(z)$ only.
After subtracting the self-energy the pion interaction energy ($V_{\Pi}$) is
\begin{eqnarray}
 V_{\Pi}  &=& \k  \left(2\pi^2\rho^2\right)^2 R^{ab}  \int d^3x dz K  \left(\dell_{i} \dell_{+a}
          H_+ \right)\left(\dell_i\dell_{-b} H_- \right)   \nn \\
          &\approx& \frac{1}{2\,3^2\pi} \frac{N_c \l \rho^4}{d^3} R^{ab}
          \left( \wh{d}_a \wh{d}_b - \frac{\d_{ab}}{3}  \right)   \ , \label{Vpi0}
\end{eqnarray}
after using $2\dell_a = - \dell_{\pm a}$ and dropping the surface terms.
$\vec{X}_+ = - \vec{X}_- = \frac{\vec{d}}{2}$ and $Z_c \approx 0$.
The matrix element of (\ref{Vpi0}) in the 2-nucleon state is
\begin{eqnarray}
\langle s_1 s_2 t_1 t_2 | V_\Pi | s_1 s_2 t_1 t_2 \rangle
&=& \frac{1}{2\, 3^5\pi} \frac{N_c \l \rho^4}{d^3}
  \left(3 (\vec{\s}_1 \cdot \wh{d} ) (\vec{\s}_2 \cdot \wh{d}) - \vec{\s}_1 \cdot \vec{\s}_2 \right)
  \left(\vec{\t}_1 \cdot \vec{\t}_2 \right) \nn \\
&\equiv& \frac{1}{4M^2}\frac{g^2_{\pi NN}}{4\pi} \inv{d^3}
  \left(3 (\vec{\s}_1 \cdot \wh{d} ) (\vec{\s}_2 \cdot \wh{d}) - \vec{\s}_1 \cdot \vec{\s}_2 \right)
  \left(\vec{\t}_1 \cdot \vec{\t}_2 \right) \label{gpin} \ ,
\end{eqnarray}
after using $\langle s_1 s_2 t_1 t_2 | R^{ab} | s_1 s_2 t_1 t_2 \rangle = \frac{1}{9} \s^a_1 \s_2^b \vec{\t}_1 \cdot \vec{\t}_2 $.
The last equality follows from the canonical $\pi\,N$ pseudovector coupling. Thus
\begin{eqnarray}
\left(\frac{g_{\pi NN}}{M}\right)^2 = \frac{8 N_c \l \rho^4}{ 3^5}
= \left(\frac{\wh{g}_A}{f_\pi}\right)^2 \ ,
\end{eqnarray}
where  $\wh{g}_A = \frac{64}{3}\k \pi \rho^2$ obtained in \cite{Hashimoto}, and $f_\pi^2 = {4\k}/{\pi}$.
This is just the Goldberger-Treiman relation following from the NN interaction using the strongly coupled
source approximation~\cite{Hashimoto}. As noted in Appendix D, our normalization of the axial-vector current
appears to be twice the normalization of the same current used in~\cite{Hashimoto}.

\subsection{Omega}

The $\w$ contribution stems from
\begin{eqnarray}
 E_{\wh{V}}  = \frac{\k}{2} \int d^3x dz K \left(\dell_z \wh{A}_0 \right)\left( \dell_z \wh{A}_0 \right) \ ,
\end{eqnarray}
where
\begin{eqnarray}
  \wh{A}_0 \approx -\frac{1}{2a\l}(G_- + G_+) \ .  \label{A0long}
\end{eqnarray}
After subtracting the self-energy the $\w$ interaction energy ($V_{\wh{V}}$) is
\begin{eqnarray}
 V_{\wh{V}}  &=& \frac{\k^2 \psi_n(Z_+)\psi_n(Z_-) m_{2n-1}^2}{(2a\l 4\pi)^2} \int d\vec{x}\
 \frac{e^{-m_{2n-1} \left(|\vec{x} - \vec{X}_- | + |\vec{x} - \vec{X}_+ | \right) }}{ |\vec{x} - \vec{X}_- ||\vec{x} - \vec{X}_+ |  }   \\
  &\approx&  \frac{N_c^2}{(8\pi)^2} \psi_n^2 m_{2n-1}^2 \int dr (4\pi r^2)\ \frac{e^{-m_{2n-1}(r+d)}}{r d}  \label{Vomegaint}\\
  &\approx&  \frac{N_c}{4}\sum_n\psi_{2n-1}^2 \frac{e^{-m_{2n-1} d}}{4\pi d} \ , \label{Vomega}
\end{eqnarray}
where we used
\begin{eqnarray}
\k \int dz  K(z) \dell_z \psi_n(z) \dell_z \psi_m(z) = m_{2n-1}^2 \d_{nm} \ .
\end{eqnarray}
The result is in agreement with (\ref{NNVV1}). At large separations, the strongly
coupled source theory and the semiclassical quantization yields the same results.
This outcome is irrespective of the use of the singular gauge (strong coupling) or
regular gauge (semiclassics). This a consequence of gauge invariance.

At short distances, gauge invariance is upset by the
delta-function singularities present in the singular gauge.
Indeed, for $\rho \ll \xi \ll 1 $ the omega contribution stems
from
\begin{eqnarray}
 E_{\wh{V}}  = \frac{\k}{2} \int d^3x dz K \left(\dell_z \wh{A}_0 \right)\left( \dell_z \wh{A}_0 \right) \ ,
\end{eqnarray}
with now $K \approx 1$ and
\begin{eqnarray}
  && \wh{A}_0 \approx -\frac{1}{2a\l}(G_-^{\mathrm{flat}} + G_+^{\mathrm{flat}}) \ ,  \label{A0mid} \\
  && G_\pm^{\mathrm{flat}} = -\frac{1}{4\pi^2}\frac{1}{\xi_\pm^2}\ ,
\end{eqnarray}
from \cite{Hashimoto}.
After subtracting the self-energy the $\w$ interaction energy ($V_{\wh{V}}'$) is
\begin{eqnarray}
 V_{\wh{V}}'  &=& \frac{\k}{(2a\l 4\pi)^2} \int d\vec{x} dz  \frac{4z^2}
 {  \left(\left(x_1-d/2\right)^2 + x_\a^2\right)^2  \left(\left(x_1+d/2\right)^2 + x_\a^2\right)^2} \nn \\
 &=&  \frac{27N_c}{2 \pi\l d^2} \int d\vec{x} dz  \frac{z^2}
 {  \left(\left(x_1-1/2\right)^2 + x_\a^2\right)^2  \left(\left(x_1+1/2\right)^2 + x_\a^2\right)^2} \nn \\
 &\approx& \frac{27\pi N_c}{4\l d^2} \ .
\end{eqnarray}
We have rescaled the variable $x_M \ra x_M/d$ in the second line and carried numerically
the integration through
\begin{eqnarray}
\int d\vec{x} dz  \frac{z^2}
 {  \left(\left(x_1-1/2\right)^2 + x_\a^2\right)^2  \left(\left(x_1+1/2\right)^2 + x_\a^2\right)^2}  \approx 4.9348 \approx \frac{\pi^2}{2} \ .
\end{eqnarray}
The omega repulsion at short distance is about  $V_{\wh{V}}/2$ in (\ref{VD2}).
The discrepancy may be due to the singularities introduced in the singular gauge and/or 
the approximation in the matching region $\rho \ll \xi \ll 1$. 
It is worth noting that the standard omega repulsion at large ditances
(\ref{Vomega}) transmutes to a 4-dimensional Coulomb repulsion in holography.

\section{Axial Form Factor}

The effective action for the SU(2)-valued axial vectors to order $\hbar^0$ follows from~\cite{FF}
\begin{eqnarray}
&& S_{\mathrm{eff}}[{\scra}^a_\m] = \sum_{b=1}^{3}
\sum_{n=1}^{\infty} \int d^4x \left[ -\frac{1}{4} \Big( \dell_\m
{a}^{b,n}_\n - \dell_\n {a}^{b,n}_\m \Big)^2 -\frac{1}{2}
m_{{a}^n}^2 ({a}^{b,n}_\m)^2
 \right. \nn \\
&& \qquad \qquad  \qquad \qquad \quad  -  \k K
{\mathbb{F}}^{b,z \m} {\scra}_\m^b
(\psi_0-\a_{a^n}\psi_{2n})
\Big|_{z=B}  \nn \\
&& \qquad \qquad  \qquad \qquad \quad  + \
 a_{{a}^n} m_{{a}^n}^2 {a}^{b,n}_\m
{\scra}^{b,\m} - \k K {\mathbb{F}}^{b,z \m} {a}_\m^{b,n} \psi_{2n}
\Big|_{z=B}  \Big]
 \ ,
\end{eqnarray}
The first line is the free action of the massive axial vector
meson which gives the meson propagator
\begin{eqnarray}
\Delta_{\m\n}^{mn,ab}(x) = \int \frac{d^4 p}{(2\pi)^4} e^{-ipx}
\left[ \frac{- g_{\m\n} - p_\m p_\n / m_{a^n}^2}{p^2 + m_{a^n}^2}
\d^{mn}\d^{ab}\right] \ ,
\end{eqnarray}
in Lorentz gauge. The rest are the coupling terms between the
source and the instanton: the second line is the direct coupling
and the last line corresponds to the
coupling mediated by the $SU(2)$ (a, a', ...) vector meson
couplings,
\begin{eqnarray}
\k K {\mathbb{F}}^{b,z \m} {a}_\m^{b,n}\psi_{2n} \ ,
\end{eqnarray}
which is large and of order $1/\sqrt{\hbar}$ since
$\psi_{2n}\sim \sqrt{\hbar}$. When $\rho$ is set to
$1/\sqrt{\l}$ after the book-keeping noted above, the coupling
scales like $\lambda\sqrt{N_c}$, or $\sqrt{N_c}$ in the large
$N_c$ limit taken first

The direct coupling drops by the sum rule
\begin{eqnarray}
\sum_{n=1}^{\infty} \a_{a^n}\psi_{2n} = \psi_0 \ = \frac{2}{\pi} \arctan z ,
\label{sumrule1}
\end{eqnarray}
following from closure in curved space
\begin{eqnarray}
\d (z-z') = \sum_{n=1}^{\infty}\ \k \psi_{n}(z)
\psi_{n}(z')K^{-1/3}(z') \ .
\label{SUM}
\end{eqnarray}
in complete analogy with VMD for the pion~\cite{Sakai1} and the electromagnetic baryon form factor~\cite{FF}.
It follows form (\ref{sumrule1}) that
\begin{eqnarray}
\sum_{n=1}^{\infty} \a_{a^n}\psi_{2n}(z_c) = \half \int_{-z_c}^{z_c} dz \dell_z \psi_0 (z)
= \frac{\k}{\pi} \int_{-z_c}^{z_c} dz  \phi_0 (z) \ .
\label{ZERO}
\end{eqnarray}
The axial vector contributions at the core sum up to the axial zero
mode.

The iso-axial current is,
\begin{eqnarray}
J^\m_{A,b}(x)  =  -\sum_{n,m}   m_{a^n}^2 a_{{a}^n} \psi_{2n} \int
d^4{y}\ \ \calq_{\n}^b(y,z) \Delta_{mn}^{\n\m}(y-x) \Big|_{z=B}
 \ ,
  \label{IVcurrent}
\end{eqnarray}
with
\begin{eqnarray}
\calq_{\n}^b(y,z) \equiv  \k  K {\mathbb{F}^b}_{z\n}({y},z) \ . \label{Q}
\end{eqnarray}
The static axial-iso-vector form factor follows readily in the form
\begin{eqnarray}
J^{bi}_{A}(\vec{q}) &=&  \int d \vec{x} e^{i\vec{q}\cdot\vec{x}} J^{bi}_{A}({x}) \nn \\
&=& (\delta^{ij} -\hat{q}^i\hat{q}^j)
\int d \vec{x} e^{i\vec{q}\cdot\vec{x}} \sum_n
\frac{\a_{{a}^n} m_{a^n}^2}{\vec{q}^{\ 2} + m_{a^n}^2}
 \psi_{2n}(z) \,\calq_j^b(\vec{x},z) \Big|_{z=B}  \ ,
\end{eqnarray}
and is explicitly transverse for massless pions. The zero momentum limit of the transverse
momentum projector is ambiguous owing to the divergence of the
spatial integrand for $\vec{q}=\vec{0}$. We use the rotationally symmetric
limit with the convention
$(\delta^{ij}-\hat{q}^i\hat{q}^j)\rightarrow  2\delta^{ij}/3$.
Thus
\begin{eqnarray}
J^{bi}_{A}(0)  =  \frac 23 \int d \vec{x} \, \k K {\mathbb{F}^b}_{zi} (\vec{x},z)\psi_0(z) \Big|_{z=B} \ ,
\end{eqnarray}
by the sum rule (\ref{ZERO}). Since the rotated instanton yields
\begin{eqnarray}
\int d\vec{x}\,  \mathbb{F}_{zi}^{R,a} =  R^{ai} \frac{4\pi^2\rho^2}{ \sqrt{z_c^2 + \rho^2} } \ ,
\end{eqnarray}
the spatial component of the axial-vector reads
\begin{eqnarray}
J^{Rbi}_{A}(0)  =  \frac{32 \k \pi \rho^2}3
(1+z_c^2)   \frac{\arctan (z_c)}{\sqrt{\rho^2 + z_c^2}}\,R^{bi} \ ,
\end{eqnarray}
In the nucleon state
\begin{eqnarray}
\langle s't'| J^{Rbi}_{A}(0)| s t \rangle
              &=&  \frac{  32 \k \pi \rho^2 (1+z_c^2) }{ 9 \sqrt{\rho^2 + z_c^2} } \arctan (z_c) (\s^b)^{ss'}  (\t^i)^{tt'} \nn \\
              &\equiv& \frac{1}{3}g_A (\s^b)^{ss'}  (\t^i)^{tt'} \ ,   \label{Jmatrix}
\end{eqnarray}
where we used $\langle s't'| R^{bi} | s t \rangle  =  - \frac{1}{3} (\s^b)^{ss'}  (\t^i)^{tt'} $. Thus
\begin{eqnarray}
  g_A = \frac{  32 \k \pi \rho^2 (1+z_c^2) }{ 3 \sqrt{\rho^2 + z_c^2} } \arctan (z_c) \approx   \frac{32}{3} \k \pi \rho^2 \ , \label{gA}
\end{eqnarray}
where the limit $\rho \ra 0$ is followed by $z_c \ra 0$.
It is $2$ times  $\wh{g}_A = \frac{16}{3} \k \pi \rho^2 $ as quoted in \cite{Hashimoto}.
This discrepancy maybe traced back to a factor of 2 discrepancy in the normalization
of the axial-vector current in \cite{Hashimoto}.

\end{document}